\documentclass[floats,floatfix,amssymb,prd,twocolumn,nofootinbib,superscriptaddress]{revtex4-2}
\usepackage[utf8]{inputenc}
\usepackage{amsmath,amssymb}
\usepackage{amsfonts}
\usepackage{bm}
\usepackage{courier}
\usepackage{graphics,orcidlink}
\usepackage{float}
\usepackage{varioref}
\usepackage{mathtools}
\usepackage[normalem]{ulem} 
\usepackage{acronym}
\usepackage{float}
\usepackage[caption=false]{subfig}
\usepackage{lipsum}
\usepackage{multirow}
\usepackage{graphicx}
\usepackage{tensor}
\usepackage{verbatim}
\usepackage{wrapfig}
\usepackage{soul} 
\allowdisplaybreaks

\def\brho{\hat{\rho}}
\def\bp{\hat{p}}
\def\bpsi{\hat{\Psi}}

\labelformat{section}{Section #1} 
\labelformat{subsubsection}{Section #1}
\labelformat{subsubsubsection}{Section #1}
\labelformat{equation}{Eq.~(#1)} 
\labelformat{figure}{Fig.~#1} 
\labelformat{subfigure}{Fig.~\thefigure#1} 
\labelformat{table}{Tab.~#1} 
\labelformat{appendix}{Appendix #1}

\begin{document}

\title{One Membrane to Love them all: Tidal deformations of compact objects from the membrane paradigm}

\author{Michela Silvestrini\orcidlink{0009-0004-2250-088X}}
\email{michela.silvestrini@unina.it}
\affiliation{Dipartimento di Fisica E. Pancini, Universit\`a di Napoli Federico II
Complesso Universitario di Monte Sant Angelo, Edificio G, Via Cinthia, I-80126, Napoli, Italy}
\affiliation{Dipartimento di Fisica, ``Sapienza'' Universit\`a di Roma \& Sezione INFN Roma1, P.A. Moro 5, 00185, Roma, Italy}

\author{Elisa Maggio\orcidlink{0000-0002-1960-8185}}
\email{elisa.maggio@aei.mpg.de}
\affiliation{Max Planck Institute for Gravitational Physics (Albert Einstein Institute), D-14476 Potsdam, Germany}

\author{Sumanta Chakraborty\orcidlink{0000-0003-3343-3227}}
\email{tpsc@iacs.res.in}
\affiliation{School of Physical Sciences, Indian Association for the Cultivation of Science, Kolkata-700032, India}

\author{Paolo Pani\orcidlink{0000-0003-4443-1761}}
\email{paolo.pani@uniroma1.it}
\affiliation{Dipartimento di Fisica, ``Sapienza'' Universit\`a di Roma \& Sezione INFN Roma1, P.A. Moro 5, 00185, Roma, Italy}

\begin{abstract}
The tidal deformability is a key observable to test the nature of compact objects in a binary coalescence. Within vacuum General Relativity, the tidal Love numbers of a four-dimensional black hole are strictly zero, while they are non-zero and model-dependent for material objects, matter distributions around black holes, or in alternative theories of gravity. Here, we develop a model-agnostic framework based on the membrane paradigm, where the tidal properties of a spherically symmetric object are encoded in the response of a fictitious membrane in terms of viscosity coefficients. We show that both neutron stars and exotic compact objects (in particular, we provide an explicit example for thin-shell gravastars) are included in this framework, provided that the bulk and shear viscosity coefficients of the membrane have a nontrivial frequency dependence. We derive a general result for the electric and magnetic tidal Love numbers for non-rotating and spherically symmetric compact objects in terms of the viscosity coefficients, discussing their behavior with the compactness of the object, and identifying the conditions for the logarithmic behavior found in previous literature for various ultracompact objects. Finally, we show that the membrane shear viscosity coefficients associated with neutron star models feature novel quasi-universal relations, which depend only mildly on the neutron-star equation of state.
\end{abstract}
\maketitle
\section{Introduction}
The tidal Love numbers~(TLNs) of a self-gravitating object measure its deformability when immersed in an external tidal field~\cite{PoissonWill}. Originally introduced to describe Earth's tidal deformations~\cite{Love1909}, TLNs also play an important role in gravitational wave~(GW) astronomy.
They affect the final stages of a binary coalescence, providing unique information about the nature of the coalescing objects.
Remarkably, within general relativity~(GR) in vacuum, the static TLNs of a four-dimensional black hole~(BH) are strictly zero~\cite{Damour_tidal, Binnington:2009bb, Damour:2009vw, Gurlebeck:2015xpa, Poisson:2014gka, Pani:2015hfa, Landry:2015zfa, LeTiec:2020bos, Chia:2020yla, LeTiec:2020spy,Bhatt:2023zsy}. This special property~\cite{Porto:2016zng, Cardoso:2017cfl} can be related to some symmetries of the perturbations of the Kerr solution in the zero-frequency limit~\cite{Hui:2020xxx, Charalambous:2021kcz, Charalambous:2021mea, Hui:2021vcv, Berens:2022ebl, BenAchour:2022uqo, Charalambous:2022rre, Katagiri:2022vyz, Ivanov:2022qqt, DeLuca:2023mio,
Gounis:2024hcm,Kehagias:2024rtz,Combaluzier-Szteinsznaider:2024sgb,
Lupsasca:2025pnt} and does not hold true for any other object (even within GR~\cite{Wade:2013hoa,Cardoso:2017cfl, Sennett:2017etc, Mendes:2016vdr, Pani:2015tga, Cardoso:2017cfl, Uchikata:2016qku, Raposo:2018rjn, Cardoso:2019rvt,Berti:2024moe}), for dynamical perturbations \cite{Bhatt:2024yyz,Katagiri:2022vyz, Ivanov:2022qqt, DeLuca:2023mio,Charalambous:2022rre}, for BHs dressed by matter distributions~\cite{DeLuca:2021ite,DeLuca:2022xlz,Katagiri:2024fpn,Chakraborty:2024gcr}, for BHs in modified gravity, as well as for asymptotically non-flat geometries~\cite{Cardoso:2017cfl, Cardoso:2018ptl, nair2024asymptotically-199, franzin2024tidal-686}, or in higher spacetime dimensions~\cite{Chakravarti:2018vlt, Chakravarti:2019aup, Cardoso:2019vof, Pereniguez:2021xcj,Charalambous:2023jgq, Rodriguez:2023xjd, Dey:2020lhq, Dey:2020pth}.

Measuring a nonzero TLN can therefore be used to exclude vacuum BHs in GR~\cite{Cardoso:2017cfl, Cardoso:2019rvt,Maselli:2018fay,Maggio:2021ans}, understand the nature of special binaries~\cite{Crescimbeni:2024cwh,Golomb:2024mmt,Crescimbeni:2024qrq}, and generically constrain the equation of state~(EoS) of nuclear matter using the GW signal from binary neutron stars~(NSs), see Refs.~\cite{GuerraChaves:2019foa,Chatziioannou:2020pqz} for some reviews.

The TLNs of compact objects other than Kerr BHs have been computed on a case-by-case basis, including boson stars~\cite{Cardoso:2017cfl, Sennett:2017etc, Mendes:2016vdr}, gravastars~\cite{Pani:2015tga, Cardoso:2017cfl, Uchikata:2016qku}, anisotropic stars~\cite{Raposo:2018rjn}, fermion-soliton stars~\cite{Berti:2024moe}, dark superfluid stars~\cite{Cipriani:2024bcc}, and other simple exotic compact objects~(ECOs)~\cite{Giudice:2016zpa, Cardoso:2019rvt} with stiff EoS at the surface~\cite{Cardoso:2017cfl}.
A recent general framework for dynamical tidal deformations of spinning compact objects~\cite{Chakraborty:2023zed} (see also \cite{Nair:2022xfm}, for the dynamical TLNs of quantum BHs) underscores the importance of taking into account the frequency dependence of the tidal response, even when eventually taking the static limit that enters at the leading post-Newtonian order in the gravitational waveforms.

In this paper, we present a novel model-agnostic approach for the TLNs of compact objects, based on the membrane paradigm. The latter was originally developed for BHs~\cite{Damour:1982,Thorne:1986iy,Price:1986yy} and more recently extended to other compact objects~\cite{Maggio:2020jml,Sherf:2021ppp,Chakraborty:2022zlq,Saketh:2024ojw} to model their ringdown.
In the standard formulation of the membrane paradigm, a static observer can replace the BH interior with a \emph{fictitious} 
membrane located at the horizon. The features of the internal spacetime are projected onto the membrane, whose physical properties are fixed by the Israel-Darmois junction 
conditions~\cite{Darmois1927,Israel:1966rt,VisserBook}. 
Likewise, the membrane paradigm for objects other than BHs replaces the object's interior with a membrane located at the object's or at some other effective radius.
The junction conditions impose that the fictitious membrane is described by a viscous fluid whose thermodynamical properties (density, pressure, and viscosity parameters) are uniquely determined if 
the membrane is demanded to act as the original object in terms of observable effects~\cite{Thorne:1986iy,Jacobson:2011dz}. 

Although our framework can in principle be applied to any theory of gravity, for simplicity we restrict to GR and spherical symmetry, for which the spacetime outside the membrane is described by the Schwarzschild metric (see~\cite{Sherf:2021ppp} for an extension to the slowly rotating case). We expect this assumption to work well also when GR is a good description near and beyond the radius of the compact object, while no specific theory is assumed to describe the object's interior. 

This paper is organized as follows.
In ~\ref{sec:membrane} we present the basic features of the membrane paradigm and how it can be used to compute the TLNs. 
In ~\ref{sec:applications} we apply the general formalism to specific examples: namely models with a generic reflectivity, NSs, and gravastars (the latter taken as an explicit example of ECOs). We will discuss several interesting findings related to the unusual properties of the fictitious membrane in these examples.
We conclude in ~\ref{sec:conclusions}, with a discussion of our main results and possible extensions. 
Further details of the computations have been presented in \ref{appbcaxial} and \ref{app:extension}.

\emph{Notations and conventions:} Throughout this work, we use geometrized $G=1=c$ units, unless otherwise mentioned. We also work with the signature convention such that flat Minkowski metric in the Cartesian coordinate system reads $\eta_{\mu \nu}=\textrm{diag}(-1,+1,+1,+1)$. 
We will use Greek indices for four-dimensional spacetime tensors, whereas we will use lower-case alphabetical indices for three-dimensional tensors.

\section{Membrane paradigm for the TLNs} \label{sec:membrane}

The membrane paradigm, originally formulated to replicate the phenomenology of BHs~\cite{Damour:1982,Thorne:1986iy,Price:1986yy}, has been recently extended to describe generic horizonless compact objects~\cite{Maggio:2020jml,Sherf:2021ppp,Chakraborty:2022zlq,Saketh:2024ojw}. The idea is to replace such an object, assumed to have mass $M$, with a fictitious membrane generated by time-like worldlines placed at a fixed radius, $R=2 M (1+\epsilon)$ (which is a spacelike hypersurface, at variance with the BH horizon), where $\epsilon$ is a positive definite quantity\footnote{We will be occasionally interested in the $\epsilon\ll1$ regime, but our formalism is general. In the following, we will provide both general results and their $\epsilon\ll1$ limit.}. The membrane must replicate the phenomenology of the compact object, thus keeping the external spacetime unchanged. For simplicity, we assume the object to be nonrotating and the membrane to be spherical (see~\cite{Saketh:2024ojw} for an extension of the membrane paradigm to the slowly rotating case). The parameter $\epsilon$, quantifying the closeness of the membrane to the would-be BH horizon, is directly linked to the compactness $\mathcal{C}=M/R$ of the object as $2\mathcal{C}=1/(1+\epsilon)$. Clearly, in the BH limit ($\epsilon\to 0$) the compactness reduces to that of a Schwarzschild BH, $\mathcal{C}=1/2$.

According to the membrane paradigm, the characteristics of the internal spacetime are projected onto the membrane, whose properties are determined by the Israel-Darmois junction conditions~\cite{Darmois1927,Israel:1966rt,VisserBook}
\begin{equation}
    [[h_{ab}]]=0\,,\;\;\;\;\;\; [[K_{ab}-h_{ab}K]]=-8\pi T_{ab}~,
    \label{condi}
\end{equation}
where $h_{ab}$ is the induced metric across the membrane, $K_{ab}$ is the extrinsic curvature, $K =K_{ab}h^{ab}$, $T_{ab}$ is the stress-energy tensor of the ($2+1$ dimensional) matter distribution on the membrane --~that is responsible for the discontinuity of the extrinsic curvature~-- and [[...]] denotes the jump of a quantity across the membrane. From the junction conditions, we will derive the boundary conditions on the spacetime perturbations outside the membrane, which in turn will determine the tidal response and the TLNs.

The junction conditions in \ref{condi} are compatible with the stress-energy tensor of a dissipative fluid,
 \begin{equation}
     T_{ab}= \rho u_a u_b +(p-\zeta \Theta) \gamma _{ab}- 2\eta \sigma_{ab}\,,
     \label{T}
 \end{equation}
where $\rho$ and $p$ are the energy density and pressure, respectively, and $u_a$ is the 3-velocity of the fluid defined in terms of its 4-velocity $U_\mu$ as $u_a=e^\mu_a U_\mu$. Here, the projector is defined as $e^\mu_a=\partial x^\mu/\partial y^a$, with $x^\mu=(t,r,\theta,\phi)$ and $y^a$ being the coordinate of the membrane (see~\cite{Maggio:2020jml} for details).
Moreover, $\Theta=D_{a}u^a$ is the expansion, $\sigma_{ab}=\frac{1}{2}(D_{c}u_{a}\gamma^c_b+D_{c}u_{b}\gamma^c_a-\Theta \gamma_{ab}$) is the shear tensor, with $\gamma_{ab}=h_{ab}+u_a u_b$ being the projection tensor and $D_{a}$ the 3-dimensional covariant derivative operator compatible with the induced metric, $h_{ab}$. Finally, the parameters $\zeta$ and $\eta$ are the bulk and the shear viscosity coefficients that govern the response of the fluid to external perturbations. 

In general, spacetime perturbations are time-dependent in the coordinate space, or, frequency-dependent when expressed in the Fourier space. In our Fourier-domain calculation we will allow the viscosity parameters $\eta$ and $\zeta$ to depend on the frequency. As we shall discuss, such frequency dependence is crucial to describe the linear response of compact objects within this framework.
It is also worth mentioning that the membrane is \emph{fictitious}, which means that its properties do not need to resemble those of ordinary or physically viable objects.
For example, within the membrane paradigm, the shear viscosity of a BH is  $\eta_{\rm{BH}}=1/(16 \pi)$, which is considered as a relatively high value, while the bulk viscosity is \emph{negative}, $\zeta_{\rm{BH}}=-1/(16\pi)$~\cite{Thorne:1986iy}.

According to the standard formulation of the membrane paradigm, we assume that the membrane is such that  $K^-_{ab}=0$. 
This assumption is justified by the fact that it is always possible to find a fictitious metric for the internal spacetime manifold of the membrane that simultaneously satisfies the junction conditions, \ref{condi}, and the requirement $K^-_{ab}=0$~\cite{Damour:1982,Thorne:1986iy,Price:1986yy}. 
We expect that different choices of the internal extrinsic curvature would correspond to different choices of the fictitious membrane and might be useful in certain contexts. We briefly discuss an extension to cases where $K^-_{ab}\neq0$, i.e., when the extrinsic curvature inside the membrane is nonvanishing in~\ref{app:extension}.

\subsection{Background spacetime}

In this section, we will describe the background spacetime and the properties of the membrane fluid associated with it. Since we are interested in static and spherically symmetric spacetimes within vacuum GR, the unique choice outside the membrane is the Schwarzschild metric, which we denote as $g^0_{\mu\nu}$, with the line element,
\begin{equation}
ds^2=-f(r) dt^2+\frac{1}{f(r)}dr^2+r^2(d\theta^2+\sin{\theta}^2 d\phi^2)\,,
\label{Sm}
\end{equation}
where $f(r)=1-2M/r$, and $M$ is the mass of the compact object. The above geometry describes the background for $r>R=2M(1+\epsilon)$, at which point the nature of spacetime changes due to the presence of the membrane fluid. Since we want to keep our analysis general, we do not specify the details of the spacetime inside the membrane; rather, the membrane itself acts as a proxy for the interior spacetime and, as we will show, the properties of the membrane depend heavily on the nature of the underlying compact object. 

Note that this formalism can be applied also when the exterior is not described by the Schwarzschild geometry (for example, due to matter fields or beyond-GR corrections) as long as the deviations from the Schwarzschild geometry decay sufficiently fast, so that one can define an \emph{effective} radius $R$ where vacuum GR equations are approximately valid.

The condition $K^{-}_{ab}=0$ is sufficient to capture the properties of the interior geometry and we will assume the same throughout the paper, except in~\ref{app:extension}, where we will discuss the consequences of choosing a nonzero value for the extrinsic curvature inside the membrane. In the above exterior geometry, the vector normal to $r=\textrm{constant}$ hypersurfaces is $n_{\mu}=(1/\sqrt{f})\delta^{r}_{\mu}$, and hence the induced metric on the membrane, located at $r=R$, reads $h_{ab}=\textrm{diag}\left(-f(R),R^{2},R^{2}\sin^2\theta\right)$, where $a\in (t,\theta,\phi)$, and the three-velocity of the membrane fluid reads $u_{a}=-\sqrt{f(R)}\delta^{t}_{a}$. Therefore, the extrinsic curvature on the membrane hypersurface has the following components, while approaching from the Schwarzschild exterior, 
\begin{equation}
K^{+}_{ab}=\textrm{diag}\left(-\frac{1}{2}\sqrt{f}f',r\sqrt{f},r\sqrt{f}\sin^2\theta \right)_{r=R}\,.
\end{equation}
The stress-energy tensor for the membrane fluid, from \ref{T}, has the following nonzero components, 
\begin{equation}
T_{ab}=\textrm{diag}\left(\rho_{0}f,r^{2}p_{0},r^{2}\sin^2\theta p_{0} \right)_{r=R}\,.
\end{equation}
Here, $\rho_{0}$ and $p_{0}$ are the unperturbed energy density and pressure of the membrane fluid, respectively. The latter can be determined by the junction condition, which relates the components of $T_{ab}$ to the components of $K^{+}_{ab}$, namely
\begin{align}
\rho_0&=-\frac{\sqrt{f(R)}}{4\pi R}~,
\label{rho0}
\\
p_0&=\frac{2 f(R)+R f'(R)}{16\pi R \sqrt{f(R)}}~.
\label{p0}
\end{align}
Note that, in the limit $R\to 2M$ (i.e., $\epsilon\to0$), the energy density identically vanishes, while the pressure diverges; this is because at the horizon it is impossible to have a static fluid element. Another interesting aspect to note is the absence of any dissipative term, such as the shear ($\eta$) and bulk ($\zeta$) viscosity coefficients, in the energy-momentum tensor of the membrane fluid for the background spacetime. Indeed, the viscosity is activated only by time-dependent perturbations of the background, which we describe below in the context of the TLNs.

\subsection{Perturbing the background spacetime: TLNs}

Here we present the dynamical equations associated with both the polar and the axial sector of metric perturbations. Subsequently, we will take the limit $\omega \to 0$ of these equations, in order to determine the static TLNs. As observed in \cite{Chakraborty:2023zed}, one should be careful with the $\omega \to 0$ limit: we will keep ${\cal O}(\omega M)$ terms throughout the computation, and we will discuss later why these terms cannot be neglected. The computation of the TLNs requires determining the perturbation equations of the exterior spacetime, and then the perturbation of the membrane to fix the boundary conditions. Therefore, as a first step, in this section we will consider time-dependent perturbations of the Schwarzschild spacetime and  express them in the Fourier space, to show explicitly the dependence on the frequency $\omega$. 

We will study separately the axial and polar perturbations, since they decouple. The axial sector depends on the metric perturbations $\delta g_{t\phi}$, and $\delta g_{r\phi}$, while the polar sector is governed by four metric perturbations: $\delta g_{tt}$, $\delta g_{rr}$, $\delta g_{tr}$ and $\delta g_{\theta \theta}$. Note that $\delta g_{\phi \phi}=\delta g_{\theta \theta}\sin^{2}\theta$, if the exterior spacetime is vacuum. By perturbing the Einstein's equations, we obtain differential equations for each of these metric perturbations. As we will demonstrate, the master equations derived for the $\delta g_{t\phi}$ and $\delta g_{tt}$ components of the metric perturbations do not depend on first order terms in $\omega$, i.e., $\mathcal{O}(M\omega)$ terms are absent in these equations. Therefore, the equations for the metric perturbations, necessary for determining the TLNs, are the same as those obtained by setting $\omega=0$ from scratch. However, the differential equations for the other components of the metric perturbations contain terms which are linear in $\omega$, and hence behave nontrivially compared to the zero-frequency case.

Specifically, we express the perturbed metric as a sum of the background metric $g^0_{\mu\nu}$, which in our case is the Schwarzschild metric, and a small perturbation $\delta g_{\mu \nu}$. In the Regge-Wheeler gauge~\cite{Regge:1957td,Zerilli:1970se}, the perturbation simplifies as the sum of an axial perturbation component, $\delta g^{\rm (ax)}_{\mu\nu}$, and a polar perturbation component, $\delta g^{\rm (pol)}_{\mu\nu}$, namely
\begin{equation}
g_{\mu\nu}=g^0_{\mu\nu}+\int d\omega e^{-i\omega t}\delta g^{\textrm{(ax)}}_{\mu\nu}+\int d\omega e^{-i\omega t}\delta g^{\textrm{(pol)}}_{\mu\nu}~,
\end{equation}
where we considered Fourier-transformed perturbations such that $\{|\delta g^{\textrm{(ax)}}_{\mu\nu}|,|\delta g^{\textrm{(pol)}}_{\mu\nu}|\}\ll |g^0_{\mu\nu}|$. The dynamical equations governing the perturbations in vacuum arise from the perturbations of the vacuum Einstein's equations, i.e., from the components of the equation $\delta R_{\mu\nu}=0$. Written explicitly, the components of the axial metric perturbations become
\begin{align}
\delta g^{\textrm{(ax)}}_{\mu\nu}=\sum_{\ell}\sin \theta \partial_{\theta}P_{\ell}(\cos \theta)
\left(
\begin{array}{cccc}
0& 0 & 0 & h_{0}\\
0 & 0 & 0 & h_{1} \\
0 & 0 & 0& 0 \\
h_{0} & h_{1} & 0 & 0\\
\end{array}
\right)~,
\end{align}
where for clarity the dependence of $h_0$ and $h_1$ on $\ell$ is left implicit. The components of the polar metric perturbations read (since we are considering a vacuum background, it follows that $(\delta g_{tt}/f)=f\delta g_{rr}\equiv H$),
\begin{align}\label{polarpert}
\delta g^{\textrm{(pol)}}_{\mu\nu}=\sum_\ell P_{\ell}(\cos \theta)
\left(
\begin{array}{cccc}
fH& H_{1}& 0 & 0\\
H_{1}& (H/f) & 0 & 0\\
0 & 0 & r^2 K & 0 \\
0 & 0 & 0 & r^2 \sin ^2 \theta  K \\
\end{array}
\right)~.
\end{align}
All entries of the above two matrices are functions of the radial coordinate alone and depend on the angular number $\ell$ (due to spherical symmetry, the perturbations do not depend on the azimuthal number $m$). 

The asymptotic behavior for the metric perturbations $h_{0}$ and $H$ at large distances reads, in the $\omega\to0$ limit,
\begin{align}
h_{0}(r)&=\frac{2S_{\ell}}{\ell r^{\ell}}+\left(\frac{2}{3\ell(\ell-1)}\right)\sqrt{\frac{2\ell+1}{4\pi}}\mathcal{B}_{\ell}r^{\ell+1}+...
\label{h0asymp}
\\
H(r)&=\frac{2M_{\ell}}{r^{\ell+1}}-\left(\frac{2}{\ell(\ell-1)}\right)\sqrt{\frac{2\ell+1}{4\pi}}\mathcal{E}_{\ell}r^{\ell}+...
\label{Hasymp}
\end{align}
where $M_\ell$ and $S_\ell$ are the mass and spin multipole moments~\cite{Cardoso:2016ryw} of the compact objects, while $\mathcal{E}_{\ell}$ and $\mathcal{B}_{\ell}$ are the electric-type and magnetic-type tidal fields, respectively. We should emphasize that both the multipole moments $M_{\ell}$ and $S_{\ell}$ as well as the tidal field $\mathcal{E}_{\ell}$ and $\mathcal{B}_{\ell}$ are accompanied by  lower multipoles having $(\ell'<\ell)$, which do not contribute to the multipolar expansion~\cite{Thorne:1980ru,Mayerson:2022ekj,Gambino:2024uge}. Given the above asymptotic expansion, we define the static TLNs as
\begin{align}
k_{\ell}^{\rm M}&=-\frac{3\ell(\ell-1)}{2(\ell+1)R^{2\ell+1}}\sqrt{\frac{4\pi}{2\ell+1}}\frac{S_{\ell}}{\mathcal{B}_{\ell}}~,
\label{tlnb}
\\
k_{\ell}^{\rm E}&=-\frac{\ell(\ell-1)}{2R^{2\ell+1}}\sqrt{\frac{4\pi}{2\ell+1}}\frac{M_{\ell}}{\mathcal{E}_{\ell}}~,
\label{tlne}
\end{align}
for the magnetic (axial) and electric (polar) sector, respectively, where $R$ is the radius of the unperturbed object.
Therefore, the determination of the TLNs can be done along the following steps --- (a) determining the differential equations for $h_{0}$ and $H$ keeping terms up to ${\cal O}(M\omega)$, (b) solving these equations, in order to compute the radial dependence of $h_{0}$ and $H$, (c) performing the asymptotic expansion of these two metric perturbations and determining the TLNs using \ref{h0asymp} and \ref{tlnb} for the magnetic part, and using \ref{Hasymp} and \ref{tlne} for the electric part. The solutions to these differential equations depend on the two arbitrary integration constants, the ratio of which is fixed by substituting appropriate boundary conditions, in turn fixing the above TLNs. Imposing the boundary conditions is where the frequency dependence of the TLN can become complicated. In what follows, we will show each of these steps, with particular emphasis on the boundary condition arising out of the membrane paradigm.

\subsubsection{Equations governing axial perturbations: Magnetic TLNs}

In the axial sector, we have two unknowns to solve for, $h_{0}$ and $h_{1}$, while there are three nontrivial Einstein's equations connecting them. It turns out that one of the equations is a consequence of the other two. In order to determine the magnetic TLN for each $\ell$, we need to eliminate $h_{1}$, in favour of $h_{0}$, using one of the two independent equations. The metric perturbation $h_{1}$ can be written in terms of $h_{0}$ and its derivative, 
\begin{align}\label{h1pfsch}
h_{1}=\frac{i\omega r^{2}\Big(rh_{0}'-2h_{0}\Big)}{r^{3}\omega^{2}-(\ell+2)(\ell-1)(r-2M)}~.
\end{align}
We can then eliminate $h_{1}$ from the other perturbed Einstein's equations and obtain the following second-order differential equation for $h_{0}$, 
\begin{widetext}
\begin{align}
h_{0}''&=\Bigg[r^{2}(r-2M)^{2}\Big\{\omega^{2}r^{3}-(\ell+2)(\ell-1)(r-2M)\Big\}\Bigg]^{-1}\Bigg\{2r^{4}(r-2M)(r-3M)\omega^{2}h_{0}'+h_{0}\Bigg[-r^{7}\omega^{4}
\nonumber
\\
&+2r^{3}(r-2M)\Big(\left\{(\ell+2)(\ell-1)-1\right\}r+4M\Big)\omega^{2}-(r-2M)^{2}(\ell+2)(\ell-1)\Big\{\ell(\ell+1)r-4M\Big\}\Bigg]\Bigg\}~.
\end{align}
\end{widetext}
Note that the above equation depends on $\omega^{2}$ and its higher powers. Since there are no terms of $\mathcal{\omega}$, we can work with the $\omega\to 0$ limit, yielding the static TLNs. Expressing the above differential equation for $h_{0}$ in the zero frequency limit, we obtain the usual one~\cite{Damour:2009vw,Binnington:2009bb} 
\begin{align}\label{diffh0genl}
\frac{h_{0}''}{h_{0}}&=\frac{\ell(\ell+1)r-4M}{r^{2}(r-2M)}~.
\end{align}
The above differential equation can be solved explicitly in terms of hypergeometric and Meijer-G functions. However, in the post-Newtonian expansion of the waveform, only the first $\ell$  multipoles of the magnetic TLNs might possibly be  relevant~\cite{Jimenez-Forteza:2018buh}, and hence we concentrate mainly on the case with $\ell=2,3$. The differential equation for $h_{0}$, associated with the $\ell=2$ mode, becomes
\begin{equation}
\dfrac{d^{2}h_{0}}{dx^{2}}=\frac{(6x-2)}{x^2(x-1)}h_{0}~,
\label{h0om0}
\end{equation}
where, we have introduced the dimensionless radial coordinate $x\equiv (r/2M)>1$. The above differential equation can be solved in terms of simple polynomials, and the general solution takes the following form, 
\begin{align}
h_{0}&=\mathcal{A}_{1}x^{2}\left(x-1\right)+\mathcal{A}_{2}\frac{12x^{3}-6x^{2}-2x-1}{3x}
\nonumber
\\
&\qquad +4\mathcal{A}_{2}x^{2}(x-1)\log\left(\frac{x-1}{x}\right)~.
\label{h0e}
\end{align}
%
where $\mathcal{A}_1$ and $\mathcal{A}_2$ are the two dimensionless arbitrary constants, to be fixed later by appropriate boundary conditions at the surface of the compact object. It is possible to determine the magnetic TLN associated with the $\ell=2$ mode in terms of the ratio of these two constants. For this purpose, we consider the limit $r\to \infty$ of \ref{h0e}, such that the perturbation to the axial part of the metric has the following asymptotic behavior,
\begin{equation}
h_{0\textrm{(asymp)}}\simeq \mathcal{A}_1 x^3\left[1 +\mathcal{O}\left(\frac{1}{x}\right)\right]+\frac{\mathcal{A}_2}{5x^2}\left[1+\mathcal{O}\left(\frac{1}{x}\right)\right]~.
\label{sv}
\end{equation}
%
A quick comparison of the above asymptotic expansion for the axial metric perturbation $h_{0}$ with \ref{h0asymp} yields $S_{2}=(\mathcal{A}_{2}/5)(2M)^{2}$ and ${\cal B}_{2}=3\mathcal{A}_{1}\sqrt{(4\pi/5)}(2M)^{-3}$. 
Substituting these results in \ref{tlnb}, we obtain
the $\ell=2$ magnetic TLN,
\begin{equation}
k^{\rm M}_{2}=-\frac{1}{15}\left(\frac{2M}{R}\right)^{5}\left(\frac{\mathcal{A}_{2}}{\mathcal{A}_{1}}\right)~.
\label{defkm}
\end{equation}
%
For completeness, we wish to compare our results with those in \cite{Damour:2009vw, Yagi:2013bca, Yagi:2013sva}. Here we define the magnetic TLNs directly from $h_{0}$, while \cite{Damour:2009vw, Yagi:2013bca, Yagi:2013sva} uses a different variable $\psi\equiv rh_{0}'-2h_{0}$, for the determination of the magnetic Love numbers. For the $\ell=2$ mode, the asymptotic expansion of $\psi$, following \cite{Damour:2009vw, Yagi:2013bca, Yagi:2013sva}, yields $\psi=\mathcal{A}_{1}^{\rm DN}(r/M)^{3}+\mathcal{A}_{2}^{\rm DN}(M/r)^{2}$. Therefore, a comparison with \ref{sv} yields the following mapping between the magnetic TLNs defined here and those in \cite{Damour:2009vw}: 
\begin{align}
k_{2}^{\rm M}=-\frac{32}{15}\left(\frac{M}{R}\right)^{5}\left(-\frac{5}{128}\frac{\mathcal{A}^{\rm DN}_{2}}{\mathcal{A}^{\rm DN}_{1}}\right)=\frac{1}{12}j_{2}~.
\end{align}
where $j_{\ell}$ are the magnetic TLNs defined in~\cite{Damour:2009vw}.
In \cite{Yagi:2013bca, Yagi:2013sva}, the magnetic TLNs are denoted by $\bar{\sigma}_{\ell}$ and, for the $\ell=2$ mode, we have the following relation:
\begin{align}\label{sigma2}
k_{2}^{\rm M}=4\left(\frac{M}{R}\right)^{5}\bar{\sigma}_{2}~.
\end{align}
Similarly, for the $\ell=3$ mode, the differential equation for $h_{0}$, as presented in \ref{diffh0genl}, can be solved analytically. This yields,
%
\begin{align}
h^{\ell=3}_{0}&=\mathcal{C}_{1}\left(\frac{x^{2}}{3}\right)\left(x-1\right)(3x-2)
\nonumber
\\
&+\mathcal{C}_{2}\left(\frac{1+5x+30x^{2}-210x^{3}+180x^{4}}{2x}\right)
\nonumber
\\
&\qquad+30\mathcal{C}_{2}x^{2}(x-1)(3x-2)\log\left(\frac{x-1}{x}\right)~.
\label{h0e3}
\end{align}
with $\mathcal{C}_{1}$ and $\mathcal{C}_{2}$ being two arbitrary constants of integration. Matching of the asymptotic expansion with \ref{h0asymp} yields, $S_{3}=(3/14)\mathcal{C}_{2}(2M)^{3}$, as well as the tidal coefficient $B_{3}=9\mathcal{C}_{1}\sqrt{(4\pi/7)}(2M)^{-4}$. Therefore, the $\ell=3$ magnetic TLN of a compact object can be expressed as,
\begin{equation}
k^{\rm M}_{3}=-\frac{3}{56}\left(\frac{2M}{R}\right)^{7}\left(\frac{\mathcal{C}_{2}}{\mathcal{C}_{1}}\right)~.
\label{defkm3}
\end{equation}
In this case, we have the following mapping between $k_{3}^{\rm M}$, defined above, and the $\ell=3$ magnetic Love number of \cite{Damour:2009vw} or the one defined in \cite{Yagi:2013bca, Yagi:2013sva}, 
\begin{align}
k_{3}^{\rm M}&=-\frac{48}{7}\left(\frac{M}{R}\right)^{7}\left(-\frac{7}{320}\frac{\mathcal{C}^{\rm DN}_{2}}{\mathcal{C}^{\rm DN}_{1}}\right)=\frac{3}{20}j_{3}\,,\\
k_{3}^{\rm M}&=\frac{45}{2}\left(\frac{M}{R}\right)^{7}\bar{\sigma}_{3}~.
\label{sigma3}
\end{align}
The arbitrary constants $(\mathcal{A}_{1},\mathcal{A}_{2})$ for the $\ell=2$ mode and $(\mathcal{C}_{1},\mathcal{C}_{2})$ for the $\ell=3$ mode crucially depend on the interior of the compact object and on the boundary conditions at its surface. For example, in the case of a Schwarzschild BH spacetime, regularity at $r=2M$ for both the perturbation and its derivative demands $\mathcal{A}_{2}=0=\mathcal{C}_{2}$\footnote{In the axial sector, the metric perturbation is well behaved at the horizon, unlike the polar sector. However, the derivative of the metric perturbation $h_{0}$ is ill-behaved at the horizon, unless $\mathcal{A}_{2}=0=\mathcal{C}_{2}$.}. This in turn implies that both the $\ell=2$ and the $\ell=3$ magnetic TLNs of a Schwarzschild BH identically vanish (the same applies to any other choices of $\ell$~\cite{Binnington:2009bb}).

\subsubsection{Equations governing polar perturbations: Electric TLNs}

In this section, we describe the electric TLNs for generic $\ell$, associated with the polar perturbations. In the case of polar perturbation, there are three nontrivial metric functions, $K$, $H_{1}$, and $H$. Among these, $H$ is directly related to the perturbation of the $g_{tt}$ component, and hence is the one we should solve for. This is why we present a differential equation for $H$ alone, expanded up to the first order in the frequency $M\omega$, which takes the following form,
\begin{align}\label{Hsole}
H=\frac{r(r-2M)\left[r(r-2M)H''+2(r-M)H'\right]}{\ell(\ell+1)r(r-2M)+4M^{2}}~.
\end{align}
Note that there are no linear-in-frequency term in the above differential equation, all the frequency dependent terms arise only at $\mathcal{O}(M^{2}\omega^{2})$. Similarly, the perturbation variable $K$ can be expressed in terms of $H$ and its derivative as,
\begin{align}\label{KHHfd}
K&=\left[\frac{(\ell+2)(\ell-1)r(r-2M)+4M(r-M)}{(\ell-1)(\ell+2)r(r-2M)}\right]H
\nonumber
\\
&\qquad +\frac{2MH'}{(\ell-1)(\ell+2)}~,
\end{align}
One can express it also in terms of $H$, $H'$ and $H''$. Alike $H$, also for the polar perturbation $K$ all the frequency dependence arises at $\mathcal{O}(M^{2}\omega^{2})$, and hence has not been reported here. Finally, $H_1$ is proportional to the frequency $\omega$, and has the following expression in terms of $H$ and $H'$,
\begin{align}\label{H1}
H_{1}&=-\left[\frac{2(\ell+2)(\ell-1)r^2 f+8(r-M)(r-3M)}{(\ell-1)\ell(\ell+1)(\ell+2)(r-2M)^{2}}\right]i\omega M H
\nonumber
\\
&\qquad -\left[\frac{2(\ell+2)(\ell-1)r^{3}f+4Mr(r-3M)}{(\ell-1)\ell(\ell+1)(\ell+2)(r-2M)}\right]i\omega H'~,
\end{align}
where we have used \ref{KHHfd}, and terms of $\mathcal{O}(M^{2}\omega^{2})$ have been ignored. Therefore, in the static limit, the polar perturbation $H_{1}$ identically vanishes, while the perturbation variable $K$ is determined in terms of $H$, which is the only polar perturbation that needs to be solved for. The differential equation for $H$, presented in \ref{Hsole}, can be solved exactly in terms of associated Legendre polynomials and yields 
\begin{align}\label{Hexterior}
H=\mathcal{P}_{1}P_{\ell}^{2}\left(1+2z\right)+\mathcal{P}_{2}Q_{\ell}^{2}\left(1+2z\right)~;
\quad
z\equiv \frac{r}{2M}-1~.
\end{align}
Here, $\mathcal{P}_{1}$ and $\mathcal{P}_{2}$ are two arbitrary constants, to be determined by applying appropriate boundary conditions at the surface of the compact object. In order to obtain the asymptotic (large $r$) behavior of the metric perturbation $H$, we can express the associated Legendre polynomials in terms of the hypergeometric functions and then obtain the following asymptotic expansion for the metric perturbation, 
\begin{align}
H_{\rm asymp}&=\mathcal{P}_{1}\frac{\Gamma(1+2\ell)}{(2M)^{\ell}\Gamma(\ell-1)\Gamma(1+\ell)}r^{\ell}
\nonumber
\\
&\quad +\mathcal{P}_{2}(2M)^{\ell+1}\frac{\sqrt{\pi}\Gamma(3+\ell)}{2^{2(\ell+1)}\Gamma(\ell+\frac{3}{2})}r^{-\ell-1}~,
\end{align}
where we have used the result that $\Gamma(-\ell)$ diverges, owing to the fact that $\ell$ is an integer. Therefore, a comparison with \ref{Hasymp} immediately yields the mass moments and the tidal moments, which for a generic $\ell$ mode read
\begin{align}
2M_{\ell}&=\mathcal{P}_{2}(2M)^{\ell+1}\frac{\sqrt{\pi}\Gamma(3+\ell)}{2^{2(\ell+1)}\Gamma(\ell+\frac{3}{2})}~;
\\
\sqrt{\frac{2\ell+1}{4\pi}}\mathcal{E}_{\ell}&=-\mathcal{P}_{1}\frac{\ell(\ell-1)\Gamma(1+2\ell)}{2(2M)^{\ell}\Gamma(\ell-1)\Gamma(1+\ell)}~.
\end{align}
The electric-type TLNs can be determined from \ref{tlne}, by simply taking the ratio of these moments with appropriate normalization factors, yielding
\begin{align}
k_{\ell}^{\rm E}&=
\frac{(\ell+2)!(\ell-2)!}{(2\ell+1)\left\{(2\ell-1)!!\right\}^{2}}\left(\frac{M}{R}\right)^{2\ell+1}\frac{\mathcal{P}_{2}}{2\mathcal{P}_{1}}~.
\end{align}
The above expression for the electric type TLNs is valid for generic $\ell$ and depends on the ratio $\mathcal{P}_{2}/\mathcal{P}_{1}$, which needs to be computed by considering perturbations in the interior and its smooth matching to the surface of the compact object. To facilitate comparison with existing literature, we relate the Love numbers defined here with those in \cite{Damour:2009vw, Yagi:2013sva}. The arbitrary constants in the expression for $H$ in \cite{Damour:2009vw, Yagi:2013sva} are defined in such a way that, asymptotically, $H\to \mathcal{P}_{1}^{\rm DN}(r/M)^{\ell}+\mathcal{P}_{2}^{\rm DN}(r/M)^{-\ell-1}$. Therefore,
\begin{align}
\frac{\mathcal{P}_{2}^{\rm DN}}{\mathcal{P}_{1}^{\rm DN}}=\left(\frac{(\ell+2)!(\ell-2)!}{(2\ell+1)\left\{(2\ell-1)!!\right\}^{2}}\right)\frac{\mathcal{P}_{2}}{\mathcal{P}_{1}}\,,
\end{align}
and hence the electric TLNs $k^{\rm DN}_{\ell}$, defined in \cite{Damour:2009vw}, and the rescaled electric TLNs $\bar{\lambda}_{\ell}$, defined in \cite{Yagi:2013bca, Yagi:2013sva}, are related to the TLNs $k_{\ell}^{\rm E}$ defined here as:
\begin{align}
k^{\rm DN}_{\ell}&=\frac{1}{2}\left(\frac{M}{R}\right)^{2\ell+1}\frac{\mathcal{P}_{2}^{\rm DN}}{\mathcal{P}_{1}^{\rm DN}}=k^{\rm E}_{\ell}~;
\\
k^{\rm E}_{\ell}&=\frac{(2\ell-1)!!}{2}\left(\frac{M}{R}\right)^{2\ell+1}\bar{\lambda}_{\ell}~.
\end{align}
The electric TLNs defined in \cite{Damour:2009vw} is identical to the ones defined here. For completeness, we also quote the result for the electric TLN associated with the $\ell=2$,
\begin{align}
k_{2}^{\rm E}&=\left(\frac{M}{R}\right)^{5}\frac{4\mathcal{P}_{2}}{15\mathcal{P}_{1}}=k_{2}^{\rm DN}=\frac{3}{2}\left(\frac{M}{R}\right)^{5}\bar{\lambda}_{2}~.
\label{defke}
\end{align}
In what follows, rather than determining the corresponding perturbations in the interior of the compact object case-by-case and then matching at the surface $r=R$, we will perturb the membrane and obtain the relevant boundary condition in a model-independent fashion. Since the membrane is designed to mimic the properties of the actual compact object, if we can determine the TLNs in terms of generic membrane quantities, this will provide a generic framework for determining the TLNs of an arbitrary compact object. To this end, we present below the perturbations of the membrane and the associated boundary conditions. 

\subsection{Perturbing the membrane}

Due to the perturbations of the spacetime metric in the exterior, the junction conditions are also perturbed. Therefore, for internal consistency, the physical quantities associated with the membrane need to be perturbed as well. Even though the dissipative part of the membrane fluid did not contribute to the background properties, the dissipative components of the membrane stress-energy tensor will play an important role in the perturbations. The pressure $p$ and the density $\rho$, and the radial coordinate $R_{\rm m}$ of the membrane are perturbed as
\begin{align}
\rho&=\rho_0 +\delta \rho (t,R,\theta)\,, 
\\
p&= p_0+ \delta p(t,R,\theta)\,, 
\\
R_{\rm m}&=R+\delta R(t,R,\theta)~.
\end{align}
Note that the perturbed density/pressure should have been evaluated at the perturbed position $R_{\rm m}$. However, the perturbation $\delta R$, in the radial position, contributes only at the second order, as we will demonstrate. Therefore, within linear perturbation theory the perturbed density/pressure depends on the unperturbed position of the membrane $R$. Since $\delta \rho(t,\theta)$, $\delta p(t,\theta)$ and $\delta R(t,\theta)$ are scalar quantities under rotation, they contribute only in the case of polar perturbations and, given the static and spherically symmetric nature of the background, these perturbations can be decomposed as follows:
\begin{eqnarray}
\delta \rho(t,R,\theta)&=&\int d\omega \sum_{\ell}\rho_1(R) P_{\ell}(\cos \theta)e^{-i\omega t}  \, ,
\\
\delta p(t,R,\theta)&=& \int d\omega \sum_{\ell}p_1(R) P_{\ell}(\cos \theta) e^{-i\omega t}\,,
\\
\delta R(t,R,\theta)&=& \int d\omega \sum_{\ell} R_1(R) P_{\ell}(\cos \theta)e^{-i\omega t}\,,
\end{eqnarray}
where $\rho_1$, $p_1$ and $R_1$ depend only on the unperturbed radius $R$ of the membrane (and on the index $\ell$, which we omit). 

For the polar sector, all perturbation variables defined above are going to contribute. As a consequence, the vector $n_{\mu}$, normal to the membrane is going to be perturbed as
\begin{align}
\delta& n_{\mu\textrm{(pol)}}=\frac{1}{\sqrt{f}}\int d\omega \sum_{\ell} e^{-i\omega t}
\nonumber
\\
&\times\left(i\omega R_{1}P_{\ell}(\cos \theta),\frac{H}{2}P_{\ell}(\cos \theta),-R_{1}\partial_{\theta}P_{\ell}(\cos \theta),0\right)\,.
\end{align}
Thus, the perturbed induced metric on the membrane has nonzero components $\delta h_{tt}$, $\delta h_{\theta \theta}$ and $\delta h_{\phi \phi}$. The three-velocity of the membrane fluid will also be perturbed, and will have nonzero components $\delta u^{t}_{\rm (pol)}$ and $\delta u^{\theta}_{\rm (pol)}$. The $\delta u^{t}$ component can be determined from the normalization condition, $u_{\mu}u^{\mu}=-1$, and reads
\begin{align}
\delta u^{t}_{\rm (pol)}=\int d\omega \sum_{\ell}\frac{e^{-i\omega t}}{2f\sqrt{f}}P_{\ell}(\cos \theta)\left(fH-f'R_{1}\right)~.
\end{align}
Given all the perturbations in the metric as well as in the matter sector associated with the membrane fluid, we can write down the perturbed components of the extrinsic curvature, as well as the components of the perturbed energy-momentum tensor. In the polar sector, for the perturbations described above, the nonzero components of the perturbed extrinsic curvature reads,
\begin{align}
\delta K^{+}_{tt\textrm{(pol)}}&=\int d\omega \sum_{\ell} \frac{e^{-i\omega t}}{4\sqrt{f}}P_{\ell}(\cos \theta)\Big[R_{1}\big(4\omega^{2}-f'^{2}
\nonumber
\\
&-2ff''\big)+2f^{2}H'+4i\omega fH_{1}+3ff'H\Big]~,
\\
\delta K^{+}_{t\theta\textrm{(pol)}}&=\int d\omega \sum_{\ell} \frac{e^{-i\omega t}}{2\sqrt{f}}\partial_{\theta}P_{\ell}(\cos \theta)\left(2i\omega R_{1}-fH_{1}\right)~,
\\
\delta K^{+}_{\theta \theta\textrm{(pol)}}&=\int d\omega \sum_{\ell} \frac{e^{-i\omega t}}{2\sqrt{f}}
\Big[R^{2}fK'+2fRK-fRH
\nonumber
\\
&+R_{1}\left(2f+Rf'-2\partial_{\theta}^{2}\right)\Big]P_{\ell}(\cos \theta)\,,
\\
\delta K^{+}_{\phi \phi\textrm{(pol)}}&=\int d\omega \sum_{\ell} \frac{e^{-i\omega t}}{2\sqrt{f}}
\Big[R^{2}fK'+2fRK-fRH
\nonumber
\\
&+R_{1}\left(2f+Rf'-2\cot \theta\partial_{\theta}\right)\Big]P_{\ell}(\cos \theta)\,.
\end{align}
These expressions have been evaluated at the unperturbed location of the membrane, i.e., at $r=R$. Moreover, $\delta K^{+}_{\rm (polar)}$ is nonzero and hence the perturbed junction conditions need to be evaluated with care. Along identical lines, the components of the perturbed energy-momentum tensor associated with the polar perturbations of the membrane fluid are given by
\begin{align}
\delta& T_{tt\textrm{(pol)}}=\int d\omega \sum_{\ell}e^{-i\omega t}P_{\ell}(\cos \theta)
\nonumber
\\
&\times \Big[f(\rho_{1}-\rho_{0}H)+\rho_{0}f'R_{1}\Big]~,
\\
\delta& T_{t\theta\textrm{(pol)}}=-R^{2}\sqrt{f}\left(\rho_{0}+p_{0}\right)\delta u^{\theta}~,
\\
\delta& T_{\theta\theta\textrm{(pol)}}=-R^{2}\Big[\left(\zeta+\eta\right)\partial_{\theta}\delta u^{\theta}+\left(\zeta-\eta\right)\cot \theta \delta u^{\theta}\Big]
\nonumber
\\
&+\int d\omega \sum_{\ell}e^{-i\omega t}P_{\ell}(\cos \theta)\Big[R^{2}\left(p_{1}+p_{0}K+\frac{i\omega \zeta}{\sqrt{f}}K\right)
\nonumber
\\
&\qquad \qquad  +R_{1}\left(2p_{0}+2\frac{i\omega \zeta}{\sqrt{f}}\right)R\Big]~,
\\
\delta& T_{\phi\phi\textrm{(pol)}}=\sin^2\theta \delta T_{\theta\theta\textrm{(pol)}}~.
\end{align}
As evident from the above components of the perturbed energy-momentum tensor in the membrane fluid, the polar perturbation brings in both the shear viscosity $\eta$ and the bulk viscosity $\zeta$. In the polar sector we have the following perturbation variables --- (a) $H$, $H_{1}$ and $K$ from the metric, (b) $\rho_{1}$, $p_{1}$, $R_{1}$ and $\delta u^{\theta}_{\rm (polar)}$ from the membrane fluid. However, as described in the previous section, in the small-frequency limit, the metric perturbations $H_{1}$ and $K$ are determined in terms of $H$. Thus, we have five perturbation variables to solve for: $\rho_{1}$, $p_{1}$, $R_{1}$, $\delta u^{\theta}_{\rm (polar)}$, and $H$. In what follows, we will use the perturbed junction conditions to determine these perturbation variables and hence determine the boundary conditions for obtaining the TLNs.  

On the other, under axial perturbations, the location of the membrane does not change; hence, the normal vector remains unchanged, i.e., $\delta n_{\mu\textrm{(ax)}}=0$. As a consequence, the induced metric is perturbed by the axial perturbation $h_{0}$, such that $\delta h_{t\phi}=\delta g_{t\phi}$ is the only nonzero component for the perturbed induced metric. Moreover, the fluid three-velocity $u^{a}$ will also be perturbed and $\delta u^{\phi}_{\rm (ax)}$ is the only perturbed component of the fluid three-velocity in the axial sector. Therefore, for axial perturbations, the only nonzero components of the perturbed extrinsic curvature are
\begin{align}
\delta &K^{+}_{t\phi\textrm{(ax)}}=\frac{1}{2}\sqrt{f}\int d\omega e^{-i\omega t} \sum_{\ell} \left(h_{0}'+i\omega h_{1}\right)
\nonumber
\\
&\qquad \qquad \qquad \times\sin \theta \partial_{\theta}P_{\ell}(\cos \theta)~,
\label{Kaxtphi}
\\
\delta &K^{+}_{\theta \phi\textrm{(ax)}}=-\frac{1}{2}\sqrt{f}\int d\omega e^{-i\omega t} \sum_{\ell} h_{1}
\nonumber
\\
&\qquad \times \left[-\cos \theta\partial_{\theta}P_{\ell}(\cos \theta)+\sin^2\theta\partial_{\theta}^{2}P_{\ell}(\cos \theta)\right]~.
\label{Kaxthetaphi}
\end{align}
It follows from the above results that $\delta K^{+}_{\rm (ax)}=0$. Similarly, the nonzero components of the perturbed membrane energy-momentum tensor, associated with the axial metric perturbations are given by
\begin{align}
\delta T_{t\phi\textrm{(ax)}}&=-\rho_{0}\int d\omega e^{-i\omega t} \sum_{\ell}h_{0}\sin \theta \partial_{\theta}P_{\ell}(\cos \theta)
\nonumber
\\
&-\sqrt{f}R^{2}\sin^2\theta \left(\rho_{0}+p_{0}\right)\delta u^{\phi}~,
\label{Taxtphi}
\\
\delta T_{\theta \phi\textrm{(ax)}}&=-\eta R^{2}\sin^2\theta \partial_{\theta}\delta u^{\phi}~,
\label{Taxthetaphi}
\end{align}
%
%
where again all the expressions are evaluated at the unperturbed location for the membrane. Notice that the shear viscosity $\eta$ appears in both the axial and polar perturbations of the membrane fluid, while the bulk viscosity arises in the polar sector only. In the axial sector, we have two perturbation variables, $h_{0}$ from the metric, and $\delta u^{\phi}_{\rm (ax)}$ from the membrane fluid, since the other metric perturbation variable $h_{1}$ is determined in terms of $h_{0}$. We will first discuss the boundary conditions and hence the TLNs in the axial sector, which is easier, before turning to the calculations in the polar sector. 

\subsubsection{Boundary condition: Axial sector}

In the axial sector, as already evident from \ref{Taxtphi} and \ref{Taxthetaphi}, the perturbations to density, pressure, and the radial coordinate of the membrane do not appear. Additionally, the bulk viscosity does not contribute to the axial perturbations of the stress-energy tensor for the membrane fluid. On the other hand, the shear viscosity $\eta$ does come into play. Since we are interested in the dynamical context, but within the small-frequency regime, we choose to expand the shear viscosity $\eta$ as a series in the frequency $\omega$, as follows:
\begin{equation}
\eta =\eta_0+i \eta_1 M\omega  +\mathcal{O}(M^2\omega^2)~,
\label{eta}
\end{equation}
where, both $\eta_{0}$ and $\eta_{1}$ are in general (complex) functions\footnote{Since the viscosity parameters are functions of $\omega$, $\epsilon$, and also $\ell$, perturbations with different frequencies and angular momenta would experience a different effective viscosity. Thus, each set of $(\epsilon, \omega, \ell)$ may be seen as defining a separate fictitious membrane.} of the compactness parameter $\epsilon$. The explicit dependence of these parameters on $\epsilon$ depend on the nature of the central compact object and we will leave it generic. Specifically, in all the expressions below we will treat $\eta_{0}=\eta_{0}(\epsilon)$ and $\eta_{1}=\eta_{1}(\epsilon)$, respectively.
For BHs, $\eta_{0}=1/16\pi$ and $\eta_{1}$ vanishes identically~\cite{Damour_viscous,Thorne:1986iy,Maggio:2020jml}. However, as we will demonstrate, for other objects it is precisely the parameter $\eta_{1}$ that plays a crucial role. By solving the perturbed junction condition (since $\delta K^{+}$ vanishes for the axial sector): 
\begin{equation}\label{pertaxial}
\delta K^{+}_{ab}-K^{+}\delta h_{ab}=-8\pi \delta T_{ab}~,
\end{equation}
associated with the axial sector, we obtain the following expression for $\delta u^{\phi}_{\rm (ax)}$,
\begin{align}\label{velpertaxial}
\delta u^{\phi}_{\rm (ax)}&=\int d\omega e^{-i\omega t}\sum_{\ell}\left(\frac{-f\left(h_{0}'+i\omega h_{1}\right)+f'h_{0}}{r\sqrt{f}\left(2f-rf'\right)}\right)_{r=R}
\nonumber
\\
&\qquad \times \textrm{cosec}\,\theta\, \partial_{\theta}P_{\ell}(\cos \theta)~,
\end{align}
and the following boundary condition for the axial perturbation $h_0$ on the membrane (for a derivation, see \ref{appbcaxial}),
\begin{align}\label{genBCaxial}
\left(\frac{h_{0}'}{h_{0}}\right)_{R}&=\frac{\gamma_{\ell}\eta_{0}f'+iM\omega\left\{\gamma_{\ell}\eta_{1}f'-\frac{rf'-2f}{8\pi M}\right\}}{\gamma_{\ell}\eta_{0}f+iM\omega \left\{\gamma_{\ell}\eta_{1}f-\frac{r(rf'-2f)}{16\pi M}\right\}}\Bigg|_{r=R}~,
\end{align}
where we have defined $\gamma_{\ell}\equiv (\ell+2)(\ell-1)$. 

As evident from the above boundary condition, for a general choice of $\eta_0$ the ratio $(h_{0}'/h_{0})$ does not depend on the viscosity at the leading order in the frequency $M\omega$, since
\begin{align}
\left(\frac{h_{0}'}{h_{0}}\right)_{R}=\frac{f'}{f}\Bigg|_{R}=\frac{2M}{R(R-2M)}+\mathcal{O}\left(M\omega\right)~.
\label{BCaxetagen}
\end{align}
Further, equating it to the ratio $(h_{0}'/h_{0})$, derived from \ref{h0e}, it follows that the ratio $(\mathcal{A}_{2}/\mathcal{A}_{1})$ must vanish, yielding a vanishing axial TLN. Thus, for any choice of nonzero $\eta_{0}$, including BHs, the axial TLNs vanish identically.

However, if we fix $\eta_0=0$, then the boundary condition depends on $\eta_{1}$ even to ${\cal O}(\omega^0)$, and takes the following form\footnote{Note that a similar behavior was discussed in~\cite{Chakraborty:2023zed}, where the case in which the reflectivity was unity was treated separately. We will explain the connection between the approach taken in~\cite{Chakraborty:2023zed} and the membrane paradigm in Sec.~\ref{sec:reflectivity}.}, 
\begin{align}
R\left(\frac{h_{0}'}{h_{0}}\right)_{R}=\frac{2\left(R^2-3MR+8\pi\gamma_{\ell}\eta_1 M^2\right)}{R^{2}-3MR+8\pi\gamma_{\ell}\eta_1 M(R-2M)}~,
\label{BCaxial}
\end{align}
where we ignored ${\cal O}(M\omega)$ terms. In terms of the dimensionless coordinate $x\equiv (r/2M)$, the above boundary condition for the axial perturbation $h_{0}$ reads, 
\begin{align}\label{BCaxialx}
x_{0}\left(\frac{1}{h_{0}}\dfrac{dh_{0}}{dx}\right)_{x_{0}}=\frac{2\left(x_{0}^2-\frac{3}{2}x_{0}+2\pi\gamma_{\ell}\eta_1\right)}{x_{0}^{2}-\frac{3}{2}x_{0}+4\pi\gamma_{\ell}\eta_1(x_{0}-1)}~,
\end{align}
where, we have defined $x_{0}\equiv (R/2M)=1+\epsilon$. Hence, given the above boundary condition, one can determine the left hand side of \ref{BCaxialx} from the exterior metric metric perturbation presented in \ref{h0e} (for $\ell=2$), and equating it to the right hand side of \ref{BCaxialx}, we obtain the following result:
\begin{align}
\frac{\mathcal{A}_{2}}{\mathcal{A}_{1}}=\frac{-3g_{2}(\epsilon)}{12g_{2}(\epsilon)\ln \left(\frac{\epsilon}{1+\epsilon}\right)+\bar{g}_{2}(\epsilon)}~,
\end{align}
where 
\begin{align}
    g_{2}(\epsilon)&=(1+\epsilon)^{2}\left(96\pi \eta_{1}\epsilon^{2}+2\epsilon^{3}+3\epsilon^{2}-1\right)\,,\\
    \bar{g}_{2}(\epsilon)&=96\pi \eta_{1}(1+2\epsilon)\left\{6\epsilon(1+\epsilon)-1\right\}+24\epsilon^{3}(3+\epsilon)\nonumber\\
    &+62\epsilon^{2}-2\epsilon-25\,.
\end{align}
Note that the shear viscosity parameter $\eta_{1}$ appearing in the above expressions is also a function of $\epsilon$. The explicit functional dependence of $\eta_{1}$ on $\epsilon$, and its $\epsilon \to 0$ limit, will be discussed in the next section. Finally, using the above result and \ref{defkm}, the magnetic TLN for $\ell=2$ mode reads
%
\begin{align}
k_{2}^{\rm M}=\frac{1}{5(1+\epsilon)^{5}}\Bigg(\frac{g_{2}(\epsilon)}{12g_{2}(\epsilon)\ln \left(\frac{\epsilon}{1+\epsilon}\right)+\bar{g}_{2}(\epsilon)}\Bigg)~.
\label{k2magnetic}
\end{align}
%
The above expression is valid for generic values of the compactness parameter $\epsilon$ and depends explicitly on $\ln \epsilon$. 

For the $\ell=3$ mode, on the other hand, the boundary condition in \ref{BCaxialx}, along with the exterior solution presented in \ref{h0e3} yields, 
\begin{align}
k_{3}^{\rm M}=\frac{3}{56(1+\epsilon)^{7}}\frac{g_{3}(x_{0})}{3\left[30g_{3}(x_{0})\log\left(\frac{\epsilon}{1+\epsilon}\right)-\bar{g}_{3}(x_{0})\right]}~,
\label{k3magnetic}
\end{align}
where, we have used \ref{defkm3}. In the above expression we have defined 
\begin{align}
g_{3}(x_{0})&=x_{0}^{5}(2x_{0}-3)(6x_{0}-5)
\nonumber
\\
&+480\pi\eta_{1}x_{0}^{3}(x_{0}-1)(2x_{0}^{2}-3x_{0}+1)~,\\
\bar{g}_{3}(x_{0})&=x_{0}\left(x_{0}-\frac{3}{2}\right)\left(3+10x_{0}+30x_{0}^{2}+120x_{0}^{3}-360x_{0}^{4}\right)
\nonumber
\\
&+240\pi\eta_{1}x_{0}\Big[1+10x_{0}-130x_{0}^{2}+240x_{0}^{3}-120x_{0}^{4}\Big]~.
\end{align}
It is worth emphasizing that the above formulas for the magnetic TLNs hold for generic choices of $\epsilon$, which is not necessarily small. In other words, they can describe both ultracompact objects with $\epsilon\ll1$ and NSs with $\epsilon={\cal O}(1)$. In the BH limit ($\epsilon \to 0$), the membrane describes an ultracompact object leading to logarithmic behavior in the magnetic TLNs, in which case we can relate the above TLNs to the ones derived using reflective boundary conditions~\cite{Cardoso:2017cfl,Chakraborty:2023zed}. We discuss each of these cases in~\ref{sec:applications}.

\subsubsection{Boundary condition: Polar sector}

In the polar sector, besides the three-vector field $u^{\mu}$, generating the membrane stress energy tensor, the energy density and pressure of the membrane fluid will also be perturbed, along with the location of the membrane. Therefore, we have five unknown functions, $\delta u^{\theta}_{\rm (pol)}$, $\rho_{1}$, $p_{1}$, $R_{1}$, and $H$. The other polar perturbations in the metric sector, namely $K$ and $H_{1}$, are determined in terms of $H$ through \ref{KHHfd} and \ref{H1}. In the polar sector, the perturbed junction conditions have contributions from both $\delta K^{+}_{ab}$, as well as $\delta K^{+}$, and hence take the following form,
\begin{align}\label{pertpolar}
\delta K^{+}_{ab}-K^{+}\delta h_{ab}-h_{ab}\delta K^{+}=-8\pi \delta T_{ab}~,
\end{align}
which provides four nontrivial equations, associated with the $(t,\theta)$, $(t,t)$, $(\theta,\theta)$ and $(\phi,\phi)$ components. The $(t,\theta)$ component of the perturbed junction condition fixes the three-velocity perturbation $\delta u^{\theta}_{\rm (pol)}$ in terms of metric perturbation and the perturbation of the location of the membrane, 
\begin{align}
\delta u^{\theta}_{\rm (pol)}=\int d\omega e^{-i\omega t}\sum_{\ell}&\partial_{\theta}P_{\ell}(\cos \theta)
\nonumber
\\
&\times \left(\frac{2i\omega R_{1}-fH_{1}}{16\pi fR^{2}(\rho_{0}+p_{0})}\right)~.
\end{align}
From the other three nontrivial equations for the polar sector, we can determine the perturbations in the density, pressure, and membrane radius, at first order in $M\omega$, assuming \ref{eta} for the small-frequency expansion of the shear viscosity, 
\begin{align}
\rho_{1}&=-\frac{\sqrt{f}}{8\pi}\left\{H'-H \frac{(R-4M)}{R(R-2 M)}\right\}+\mathcal{O}(\omega^2)~,\label{eq:rho1}
\\
p_{1}&=\frac{i\omega \zeta R(R-2M)H'}{\sqrt{f}(R-3 M)}-\frac{H}{16\pi\sqrt{f}}\Bigg[\frac{R-3M}{R^{2}}
\nonumber
\\
&\quad +16i\omega\pi\zeta\left(\frac{R-4M}{R-3M}\right)\Bigg]+\mathcal{O}(\omega^2)~,\label{eq:p1}
\\
R_1&=\frac{16\pi R\eta fH_{1}}{2f-Rf'-32\pi i \eta \omega R},\label{eq:R1}
\end{align}
where all the quantities have been evaluated at $r=R$. To derive the boundary condition for the polar perturbation $H$ at $r=R$, we notice that $p_{1}$ and $\rho_{1}$ are not independent, but are rather related to each other through the barotropic EoS, $p=p(\rho)$,
\begin{equation}\label{EoSmp}
p_{1}=\left(\frac{\partial p_{0}}{\partial \rho_{0}}\right)\rho_{1} +\left(\frac{\partial p_{0}}{\partial R}\right)R_{1}=c_{\rm s}^{2}\rho_{1}+\mathcal{O}(\omega)~,
\end{equation}
where $c_{s}=\sqrt{\partial p_0/\partial \rho_0}$ is the speed of sound through the membrane fluid and we used the fact that $R_1\propto H_1\propto\omega$. Solving for $(H'/H)_R$, assuming $\eta_0=0$ (so that $\eta={\cal O}(M\omega)$), and ignoring ${\cal O}(M^{2}\omega^2)$ terms, the boundary condition for the polar metric perturbation $H$ reads 
\begin{widetext}
\begin{align}
\left(\frac{H'}{H}\right)_{R}&=\frac{M\left(R^{2}-6MR+6M^{2}\right)-16\pi i\omega \zeta R^{2}(R-4M)(R-2M)}{R(R-2M)\left[\left(R^{2}-3MR+3M^{2}\right)-16\pi i\omega\zeta R^{2}(R-2M)\right]}~.
\label{BCH}
\end{align}
\end{widetext}
%
Note that, to ${\cal O}(M\omega)$ and when $\eta_0=0$, the above boundary condition is independent of $\eta$.
In the $\omega \to 0$ limit, for generic values of the bulk viscosity $\zeta$, the above boundary condition for the polar metric perturbation $H$ becomes independent of $\zeta$, taking the following form
\begin{align}\label{regularzeta}
\left(\frac{H'}{H}\right)_{R}=-\frac{M \left(6 M^2-6 M R+R^2\right)}{R (2 M-R) \left(3 M^2-3 M R+R^2\right)}~.
\end{align}
In general, the bulk viscosity enters the boundary condition through the combination $\omega \zeta$. Hence, if $\zeta$ is regular in the zero-frequency limit, the boundary condition in the polar sector is independent of the viscosity. As evident from \ref{regularzeta}, in the case of $\zeta$ being regular in the zero-frequency limit, it follows that $(H'/H)$ diverges in the $R\to 2M$ limit, yielding vanishing polar TLNs. The only way to preserve the dependence of the boundary condition in the polar sector (and hence of the polar TLNs) on the bulk viscosity, $\zeta$, necessary to obtain non-zero polar TLNs, is to assume an inverse dependence of $\zeta$ on the frequency. Therefore, we choose to expand $\zeta$ as 
\begin{equation}
\zeta =i\frac{\zeta_{-1}}{M\omega}+\zeta_0+{\cal O}(M\omega)~.
\label{zeta}
\end{equation}
Alike the shear viscosity, the parameters $\zeta_{-1}$ and $\zeta_{0}$ are also (complex) functions of $\epsilon$, with their explicit dependence being dictated by the nature of the central compact object under consideration. In the following we will focus on the frequency dependence of the TLNs, while subsequent sections will provide a discussion of the $\epsilon$-dependence. Note that including a term $\zeta_{-2}/\omega^{2}$ in \ref{zeta} does not modify any of these results, as the boundary conditions would demand $\zeta_{-2}=0$.

While the fact that the bulk viscosity diverges in the static limit might appear worrisome, but we remind ourselves that the membrane itself is \emph{fictitious} so it does not need to reproduce the properties expected for a well-behaved physical system. For example, the BH case corresponds to a \emph{negative} bulk viscosity, which is related to the acausal nature of the membrane fluid if the latter needs to reproduce the global properties of a BH spacetime~\cite{Thorne:1986iy}. Indeed, we will explicitly show that mapping this framework to concrete models such as NSs and ECOs requires $\zeta_{-1}\neq0$. With the above expansion, the boundary condition for the polar metric perturbation $H$ in the static limit reads
\begin{widetext}
\begin{align}
\left(\frac{H'}{H}\right)_{R}&
=-\frac{M(R-3M)^2}{MR\left(3 M^{2}-3MR+R^{2}\right)+16\pi\zeta_{-1}R^{3}(R-2M)}-\frac{2 M}{R(R-2 M)}+\frac{1}{R}~,
\label{BCHn}
\end{align}
%
which is valid for any value of the angular number $\ell$. As evident the polar boundary condition, with $\eta_{0}=0$, does not depend on the shear viscosity, but depends explicitly on $\zeta_{-1}$ in the static limit. Therefore, the linear-in-frequency term in the shear viscosity is uniquely determined from the axial boundary condition, while the $\zeta_{-1}$ term in bulk viscosity is determined uniquely from the polar boundary condition.

Given the exterior solution to the metric perturbation $H$, as presented in \ref{Hexterior}, from the boundary condition in \ref{BCHn}, the electric TLN for the $\ell=2$ mode reads,
%
\begin{align}\label{k2E}
k_{2}^{\rm E}=\frac{A(\epsilon) }{5(1+\epsilon)^4\left[B(\epsilon)+C(\epsilon) \log \left(\frac{1}{\epsilon }+1\right)\right]}~,
\end{align}
where we have introduced three functions of the compactness parameter $\epsilon$ and bulk viscosity parameter $\zeta_{-1}$, namely 
%
\begin{eqnarray}
A(\epsilon) &=&\epsilon \left[64 \pi \zeta_{-1} \epsilon^4+4 (64 \pi \zeta_{-1}+1) \epsilon^3+(320 \pi \zeta_{-1}+3) \epsilon^2+(128 \pi \zeta_{-1}+3) \epsilon+1\right] \,, 
\nonumber 
\\
B(\epsilon) &=&768 \pi \zeta_{-1} \epsilon^5+48 (72 \pi \zeta_{-1}+1) \epsilon^4+(5248 \pi \zeta_{-1}+60) \epsilon^3+(3008 \pi \zeta_{-1}+46) \epsilon^2+28 (16 \pi \zeta_{-1}+1) \epsilon+3\,, 
\nonumber 
\\
C(\epsilon) &=&-12 \epsilon (\epsilon+1) \left[64 \pi \zeta_{-1} \epsilon^4+4 (64 \pi \zeta_{-1}+1) \epsilon^3+(320 \pi \zeta_{-1}+3) \epsilon^2+(128 \pi \zeta_{-1}+3) \epsilon+1\right]~.
\end{eqnarray}
Note that $A(\epsilon)\sim \epsilon$, $B(\epsilon)\sim (3+28\times 16\pi \zeta_{-1}\epsilon)+28\epsilon$, and $C(\epsilon)\sim -12\epsilon$, to leading order in the compactness parameter $\epsilon$. Therefore, unless $\zeta_{-1}\epsilon=-3/(28\times 16\pi)+\mathcal{O}(\epsilon)$, the TLN associated with $\ell=2$ mode vanishes identically as $\epsilon\to0$. Thus, in order to get nonvanishing TLNs as $\epsilon\to0$, one needs the bulk viscosity to be inversely dependent on the frequency and also inversely (and fine-tuned) dependent on the compactness parameter $\epsilon$, i.e. $\zeta\sim (\omega\epsilon)^{-1}$ as $\omega\sim0$ and $\epsilon\sim0$. Thus, just as for BHs the bulk viscosity is negative, leading to teleological behaviour, for ECOs also the bulk viscosity has a inverse-in-frequency and inverse-in-compactness behaviour, suggesting that in both the cases, the membrane is fictitious.

Along identical lines, one can determine the solution for the metric perturbation $H(r)$ with $\ell=3$ in the exterior of the compact object by substituting $\ell=3$ in \ref{Hexterior}. Matching the same with the boundary condition presented above, the octupolar ($\ell=3$) electric TLN reads 
\begin{align}\label{k3E}
k_{3}^{\rm E}=\frac{D(\epsilon) }{56 (\epsilon +1)^6 \left[E(\epsilon)+F(\epsilon) \log \left(\frac{\epsilon}{1+\epsilon}\right)\right]}~,
\end{align}
where we defined
%
\begin{eqnarray}
D(\epsilon) &=&\epsilon  \left[256 \pi  \zeta_{-1} \epsilon ^5+12 (80 \pi  \zeta_{-1}+1) \epsilon ^4+16 (80 \pi  \zeta_{-1}+1) \epsilon ^3+4 (176 \pi  \zeta_{-1}+3) \epsilon ^2+2 (64 \pi  \zeta_{-1}+3) \epsilon +1\right]\,, 
\nonumber 
\\
E(\epsilon) &=&7680 \pi  \zeta_{-1} \epsilon ^6+120 (272 \pi  \zeta_{-1}+3) \epsilon ^5+20 (2576 \pi  \zeta_{-1}+33) \epsilon ^4+20 (1808 \pi  \zeta_{-1}+27) \epsilon ^3+2 (5008 \pi  \zeta_{-1}+155) \epsilon ^2
\nonumber
\\
&&+(416 \pi  \zeta_{-1}+82) \epsilon +3\,, 
\nonumber 
\\
F(\epsilon) &=&30 \epsilon  (\epsilon +1) \left[256 \pi  \zeta_{-1} \epsilon ^5+12 (80 \pi  \zeta_{-1}+1) \epsilon ^4+16 (80 \pi  \zeta_{-1}+1) \epsilon ^3+4 (176 \pi  \zeta_{-1}+3) \epsilon ^2+2 (64 \pi  \zeta_{-1}+3) \epsilon +1\right]~.\nonumber\\
\end{eqnarray}
Also in this case $\zeta_{-1}$ must scale as $\epsilon^{-1}$ for the $\ell=3$ TLN to be nonzero. However, the coefficient of the $\epsilon^{-1}$ term is different between the $\ell=2$ and $\ell=3$ cases, suggesting that $\zeta_{-1}$ depends on the angular number $\ell$ as well. This can be seen from the TLN for generic $\ell$. Using the solution for the metric perturbation $H(r)$ in the exterior for generic values of $\ell$, one can determine the electric type TLNs for generic $\ell$, which reads,
\begin{align}
k_{\ell}^{\rm E}=-\frac{2^{-4 \ell-3}\pi(\epsilon +1)^{-2 \ell-1} \Gamma (\ell-1) \Gamma (\ell+3)}{\Gamma \left(\ell+\frac{1}{2}\right) \Gamma \left(\ell+\frac{3}{2}\right)} \frac{f(\epsilon,\ell,\zeta_{-1})}{g(\epsilon,\ell,\zeta_{-1})}\,.
\end{align}
The two functions $f(\epsilon,\ell,\zeta_{-1})$ and $g(\epsilon,\ell,\zeta_{-1})$, dependent on the compactness parameter $\epsilon$, as well as the angular number $\ell$ and bulk viscosity parameter $\zeta_{-1}$, appearing in the TLN for generic values of $\ell$, has the following expressions,
%
\begin{align}
f(\epsilon,\ell,\zeta_{-1})&=\Big[-4\epsilon  \left[32 \pi \zeta_{-1} (4\epsilon -1) (\epsilon +1)^2+\epsilon  (2\epsilon +3)\right]-\ell (2\epsilon +1) \big\{2\epsilon  \left[64 \pi \zeta_{-1} (\epsilon +1)^2+2\epsilon +1\right]+1\big\}+1\Big]P_\ell^2(2\epsilon +1)
\nonumber
\\
&\quad+(\ell-1) \left[2\epsilon  \left(64 \pi \zeta_{-1} (\epsilon +1)^2+2\epsilon +1\right)+1\right]P_{\ell+1}^2(2\epsilon +1)\,, 
\nonumber 
\\
g(\epsilon,\ell,\zeta_{-1})&=\Big[-4\epsilon  \left[32 \pi \zeta_{-1} (4\epsilon -1) (\epsilon +1)^2+\epsilon  (2\epsilon +3)\right]-\ell (2\epsilon +1) \big\{2\epsilon  \left[64 \pi \zeta_{-1} (\epsilon +1)^2+2\epsilon +1\right]+1\big\}+1\Big] Q_\ell^2(2\epsilon +1)
\nonumber
\\
&\quad+(\ell-1) \left[2\epsilon  \left[64 \pi \zeta_{-1} (\epsilon +1)^2+2\epsilon +1\right]+1\right]  Q_{\ell+1}^2(2 \epsilon +1)\,.
\end{align}
\end{widetext}
The above expressions are valid for generic choices of $\epsilon$, not necessarily small. For ultracompact objects, i.e., for small $\epsilon$, the electric TLNs generically display a logarithmic behavior (for generic $\ell$, the logarithmic behavior originates from the associated Legendre polynomial $Q^{2}_{\ell+1}$), although this also depends on the form of $\zeta_{-1}(\epsilon)$, as discussed before and also elaborated in the next section.

\section{Applications to compact objects}\label{sec:applications}

In this section, we discuss some applications of the membrane paradigm for determining the TLNs of certain class of compact objects. In particular, we shall show that our framework can accommodate known particular cases by connecting them with the parameters of the membrane fluid and hence determining the associated TLNs from the results derived in the previous section. We also describe the $\epsilon \to 0$ limit of the shear and bulk viscosity parameters described in the previous section. 

\subsection{Connecting reflectivity and viscosity}
\label{sec:reflectivity}

The TLNs can be determined in terms of the effective reflectivity of the compact object and its compactness. Earlier results existed for static TLNs with unit reflectivity~\cite{Cardoso:2017cfl}. Recently, Ref.~\cite{Chakraborty:2023zed} (see also \cite{Nair:2022xfm}) developed a framework to compute the tidal response of a Kerr-like compact object in terms of its reflectivity, compactness, and spin, in a dynamical context. Assuming zero spin, we now show that the membrane paradigm can also be directly mapped to that framework, and hence we obtain a one-to-one correspondence between the Detweiler reflectivity, defined in~\cite{Chakraborty:2023zed}, and the viscosity parameters $\eta_{1}$ and $\zeta_{-1}$.

For this purpose, we consider gravitational perturbations encoded in the Detweiler master function $\psi(r_{*})$ of the background spacetime~\cite{Chakraborty:2023zed}, with $r_{*}(r)$ being the tortoise coordinate. The gravitational perturbation is partially absorbed and partially reflected at the boundary of the membrane, located at $r_{*}^{0}=r_{*}(R)$. Since the perturbation $\psi(r_{*})$ satisfies a Schr\"{o}dinger-like equation, with an effective potential, which vanishes near the BH horizon, it follows that for $R\approx 2M$, i.e., for $\epsilon \ll 1$, the perturbation takes the following form,
\begin{equation}
\psi(r_{*})\simeq {\cal R} e^{i \omega (r_{*}-r_{*}^{0})}+e^{-i \omega (r_{*}-r_{*}^{0})}~, 
\label{psi}
\end{equation}
with $\mathcal{R}$ being the reflectivity of the membrane. It is worth stressing that the above behavior is valid only for membranes located near the would be BH horizon ($\epsilon\ll1$), as the effective potential vanishes in this case. 

In order to match the results derived here with those in~\cite{Chakraborty:2023zed}, we need to relate the Regge-Wheeler and the Zerilli functions to the Detweiler function in the zero-rotation limit. This can be achieved by decomposing the Detweiler function in terms of the Regge-Wheeler and the Zerilli functions in the zero rotation limit, in which one schematically obtains
\begin{align}
X^{\rm axial}_{\ell m}&=A^{\rm RW}_{\ell m}(r)\Psi^{\ell m}_{\rm RW}+B_{\ell m}^{\rm RW}(r)\left(\dfrac{d\Psi_{\rm RW}^{\ell m}}{dr}\right)~,
\label{axial}
\\
X^{\rm polar}_{\ell m}&=A^{\rm Z}_{\ell m}(r)\Psi^{\ell m}_{\rm Z}+B_{\ell m}^{\rm Z}(r)\left(\dfrac{d\Psi_{\rm Z}^{\ell m}}{dr}\right)~,
\label{polar}
\end{align}
where, $X_{\ell m}^{\rm axial/polar}$ are the axial and polar parts of the Detweiler function, and the explicit radial dependence of the functions $A^{\rm RW/Z}_{\ell m}(r)$ and $B_{\ell m}^{\rm RW/Z}(r)$ can be found in Appendix~D of \cite{Chakraborty:2023zed}. 
In turn, this implies that the Detweiler reflectivity can be decomposed into an axial and a polar part, such that $\mathcal{R}=F(\mathcal{R}_{\rm ax},\mathcal{R}_{\rm pol})$, where $F$ is some function, the specific form of which will not be relevant for our purpose here. On the other hand, the Regge-Wheeler and the Zerilli functions are related to the respective viscosity parameters of the membrane fluid through the following boundary conditions --- (a) in the axial case, we can express the boundary condition for the Regge-Wheeler function, $\psi_{\rm RW}$, as follows~\cite{Maggio:2020jml},
\begin{equation}
\left(\frac{1}{\psi_{\rm RW}}\dfrac{d\psi_{\rm RW}}{dr_{*}}\right)_{R}=-\frac{i\omega }{16\pi\eta}-\frac{R^2}{2(R-3 M)}V_{\rm RW}(R)~,
\label{bcax}
\end{equation}
with the effective potential being given by
\begin{equation}
V_{\rm RW}(r)=\left(1-\frac{2M}{r}\right)\left(\frac{\ell (\ell+1)}{r^2}-\frac{6M}{r^3}\right)~.
\end{equation}
(b) In the polar case, the Zerilli function $\psi_{\rm Z}$, satisfies the following boundary condition~\cite{Maggio:2020jml}
\begin{equation}
\frac{1}{\psi_{\rm Z}}\left(\dfrac{d\psi_{\rm Z}}{dr_{*}}\right)=-16 \pi i\eta\omega+\frac{G(R,\omega,\eta,\zeta,\ell)}{M}~,
\label{bcpol}
\end{equation}
with $G(R,\omega,\eta,\zeta,\ell)$ being a complicated function that can be found in Appendix~A of \cite{Maggio:2020jml}. Relating the Regge-Wheeler and the Zerilli functions to Detweiler function, through \ref{axial} and \ref{polar}, yields the desired relations between the shear and the bulk viscosity parameter $(\eta,\zeta)$ and the corresponding reflectivities $\mathcal{R}_{\rm ax}$ and $\mathcal{R}_{\rm pol}$. 

Note that even though for the exterior vacuum geometry the axial and the polar perturbations are related by the Chandrasekhar transformation, the boundary conditions on the membrane in the axial and the polar sector are independent of one another. This is because the Chandrasekhar transformation holds only in vacuum, while the membrane provides an effective description of the compact object, and hence depicts nonvacuum spacetime. Thus, the axial and polar perturbations inside the compact object are not related to one another, which in turn manifests itself with different boundary conditions for the polar and the axial sector at the membrane. Hence, the shear viscosity $\eta$ and the bulk viscosity $\zeta$ are independent of one another. This also follows from the fact that the two reflectivities $\mathcal{R}_{\rm ax}$ and $\mathcal{R}_{\rm pol}$ are independent.

Moreover, the axial boundary condition for an ECO, whose surface is close to that of the horizon, does not depend on the angular number $\ell$, since the effective potential in the axial sector vanishes in the near-horizon limit. Thus, we expect the connection between shear viscosity $\eta$ and the Detweiler reflectivity $\mathcal{R}_{\rm ax}$ to be independent of the angular number $\ell$. Indeed, in the axial sector, for $\ell$=2 and $\epsilon\ll1$, we obtain the following relation between the shear viscosity parameter $\eta(\omega)$ and the Detweiler reflectivity ${\cal R}_{\rm ax}(\omega)$:
\begin{equation}
    \eta=\frac{1}{16\pi}\times\frac{4M\omega\left(1+\mathcal{R}_{\rm ax}\right)+i\left(1-\mathcal{R}_{\rm ax}\right)}{4M\omega\left(1-\mathcal{R}_{\rm ax}\right)+i\left(1+\mathcal{R}_{\rm ax}\right)}\,.
\label{etaR}
\end{equation}
One can verify that the above relation is valid for generic choices of $\ell$, as expected.
Subsequently, expanding both the Detweiler reflectivities $\mathcal{R}_{\rm ax}$ and $\mathcal{R}_{\rm pol}$ for small frequencies in the same way as~\cite{Chakraborty:2023zed}:
\begin{equation}
{\mathcal R}(\omega) = {\mathcal R}_0+ iM\omega {\mathcal R}_1 +{\mathcal O}(M^2\omega^2) \,,
\label{reflectivity}
\end{equation}
and comparing with the small-frequency expansion of the shear viscosity, presented in \ref{eta} (in the limit $\epsilon\to0$), we obtain the mapping in the small-frequency expansion:
\begin{equation}
\eta_{0}=\frac{1}{16\pi}\left(\frac{1-\mathcal{R}^{\rm ax}_{0}}{1+\mathcal{R}^{\rm ax}_{0}}\right)\,.
\end{equation}
As expected, the BH case (perfect absorption and hence ${\cal R}^{\rm ax}_{0}=0$) corresponds to $\eta_0=1/(16\pi)$~\cite{Thorne:1986iy}. On the other hand, if ${\cal R}^{\rm ax}_{0}=1$, we obtain $\eta_0=0$. This is precisely the scenario yielding nonzero TLNs for the compact object in the zero-frequency limit. Thus, once again, in agreement with~\cite{Chakraborty:2023zed}, we find that nonzero TLNs in the zero-frequency limit require $\mathcal{R}^{\rm ax}_{0}=1$ or, equivalently, $\eta_{0}=0$. This is consistent with our findings in the present work, though from a completely different point of view. 

For the relevant case $\eta_0=0$ (that is, ${\cal R}^{\rm ax}_0=0$), we also find a relation between $\eta_1$ and ${\cal R}^{\rm ax}_1$:
\begin{equation}
\eta_1=-\frac{\mathcal{R}^{\rm ax}_1+8}{32\pi}~,
\label{eta1}
\end{equation}
which is also independent of the angular number $\ell$. Thus, a compact object will have nonzero TLNs if and only if its zero-frequency Detweiler reflectivity is unity and, within the membrane paradigm, the shear viscosity is proportional to the frequency $\omega$. The above holds for the axial case, we will now show that a similar result exists also for the polar sector.  

In the polar sector, relating the Zerilli function with the Detweiler function yields the following relation between the bulk viscosity $\zeta(\epsilon,\omega)$ and the Detweiler reflectivity ${\cal R}_{\rm pol}(\omega)$, in the small-$\epsilon$ limit, 
%
\begin{align}
\zeta^{\ell=2}_{-1}
=-\frac{3}{448\pi \epsilon}-\frac{3\left(\mathcal{R}^{\rm pol}_1+18\right)}{3136\pi}~.
\label{zetam1l2}
\end{align}
and,
%
\begin{align}
\zeta^{\ell=3}_{-1}  
=-\frac{3}{416\pi \epsilon}-\frac{3\left(5\mathcal{R}^{\rm pol}_1+59\right)}{10816\pi}~,
\end{align}
for $\ell=2$ and $\ell=3$, respectively. As evident, unlike the axial case, the connection between bulk viscosity and the corresponding polar reflectivity depends on the angular number $\ell$. Further, the above relation assumes that $\mathcal{R}_{0}^{\rm pol}=1$.
As expected, the leading-order behavior is $\zeta_{-1}^{\ell}\sim1/\epsilon$, with the right coefficient in order to obtain a nonvanishing TLNs.

Finally, let's compute the magnetic and electric TLNs in terms of the reflectivity, using the expressions derived above, in terms of $\eta_{1}$ and $\zeta_{-1}$, through the membrane paradigm. This is achieved by simply substituting the relations found above between $\eta_1$ or $\zeta_{-1}$ and the corresponding reflectivities in the expressions for TLNs derived earlier. In particular, by plugging \ref{eta1} into the expression presented in \ref{k2magnetic}, for the $\ell=2$ magnetic TLN we find 
\begin{align}
k_{2}^{\rm M}&=\frac{1 \text{}}{60 \log \left(\epsilon\right)-15 \mathcal{R}^{\rm ax}_1+5}
\nonumber
\\
&\quad+\frac{\epsilon\left[-3\mathcal{R}^{\rm ax}_1-60 \log (\epsilon )-89\right]}{5 \left[3\mathcal{R}^{\rm ax}_1-12 \log (\epsilon )-1\right]^2}+\mathcal{O}(\epsilon^2)\,.
\label{tlnax}
\end{align}
The leading order terms, namely the coefficient of $\log(\epsilon)$ as well as that of the reflectivity $\mathcal{R}_{1}^{\rm ax}$, coincide with the result of \cite{Chakraborty:2023zed}. Hence the membrane paradigm provides consistent expressions of the TLNs of an ECO with reflectivity. Likewise, we obtain the following TLNs associated with the $\ell=2$ and $\ell=3$ modes in the polar sector: 
\begin{align}
k_{2}^{\rm E}&=\frac{1 \text{}}{60 \log \left(\epsilon\right)-15 \mathcal{R}^{\rm pol}_1+5}
\nonumber
\\
&\quad+\frac{\text{}\epsilon\left(18(\mathcal{R}^{\rm pol}_1)^{2}+417 \mathcal{R}^{\rm pol}_1-420 \log (\epsilon )+1996\right)}{35\left(3 \mathcal{R}^{\rm pol}_1-12 \log (\epsilon )-1\right)^2}\nonumber
\\
&\quad+\mathcal{O}(\epsilon^2)~,
\label{kER}
\\
k_{3}^{\rm E}&=\frac{1}{28\left(60 \log (\epsilon )-15\mathcal{R}^{\rm pol}_{1}+77
\right)}+\mathcal{O}(\epsilon)~,\label{kER3}
\end{align}
which also agree at the leading order in $\epsilon$ with the result of~\cite{Chakraborty:2023zed}. 

Interestingly, if $\mathcal{R}_{1}^{\rm ax}=\mathcal{R}_{1}^{\rm pol}$, it follows that $k_{2}^{\rm M}=k_{2}^{\rm E}$, i.e., the magnetic and electric TLNs are the same, to leading order in the $\epsilon\to0$ limit, while the ${\cal O}(\epsilon)$ terms are different. 
This is valid also for $\ell=3$; indeed 
\begin{align}
k_{3}^{\rm M}&=\frac{1}{28\left[60 \log (\epsilon )-15\mathcal{R}^{\rm pol}_{1}+77\right]}+\mathcal{O}(\epsilon)~,
\end{align}
which agrees with~\ref{kER3} if $\mathcal{R}_{1}^{\rm ax}=\mathcal{R}_{1}^{\rm pol}$. Unfortunately, due to the nature of the equations governing the magnetic sector we cannot test this equivalence for generic $\ell$, but we suspect it is always valid.

This general result agrees with and generalizes some particular cases~\cite{Cardoso:2017cfl,HuiSantoni}, which where obtained by imposing $\mathcal{R}^{\rm ax}=1=\mathcal{R}^{\rm pol}$ at any frequency, corresponding to setting the linear-in-frequency components $\mathcal{R}_{1}^{\rm ax}=\mathcal{R}_{1}^{\rm pol}=0$. However, in general, the axial and the polar sectors might have different reflectivity, and hence the electric and magnetic TLNs are generically different.

\subsection{TLNs of NSs from membrane paradigm}

In the previous section, we have related the reflectivity of any sufficiently compact object with the parameters of the membrane paradigm, in particular with the shear and bulk viscosity. That connection is only possible when $\epsilon\to0$, since this condition is required to express the boundary conditions in terms of incident and reflected waves.
To illustrate the generality of our framework, we now apply it to a concrete and physically relevant example, namely perfect-fluid NSs, for which $\epsilon$ is not arbitrarily small (in fact, for typical EoS $\epsilon={\cal O}(1)$ near the maximum mass). In this case, we will also unveil novel quasi-universal relations for some of the viscosity coefficients that are only mildly sensitive to the EoS of the nuclear matter. We start by reviewing the background spacetime.

In the static case, the spherically symmetric exterior geometry is described by the Schwarzschild metric, as in \ref{Sm}, while the spacetime inside the star, being also spherically symmetric and static, has the following line element
\begin{equation}
ds^{2}=-e^{\nu}dt^{2}+e^{\lambda}dr^{2}+r^{2}d\Omega^{2}\,,
\end{equation}
where $\nu$ and $\lambda$ are functions of the radial coordinate $r$ alone. Further, we assume that the interior of the NS consists of an isotropic perfect fluid\footnote{For the NS fluid we use hatted quantities to distinguish them from those of the fictitious membrane previously introduced.} with energy density $\brho$ and pressure $\bp$, such that the associated stress-energy tensor becomes,
\begin{equation}
T_{\mu \nu}^{\rm (ns)}=(\brho+\bp)\hat{u}_{\mu}\hat{u}_{\nu}+\bp g_{\mu \nu} \,,
\end{equation}
where $\hat{u}^{\mu}$ is the four-velocity of the fluid inside the NS. Expressing, $e^{-\lambda}=1-2m(r)/r$, where $m(r)$ is the mass function, the background Einstein's equations reduce to the well-known Tolman-Oppenheimer-Volkoff equations,
\begin{eqnarray}
m'(r) &=& 4 \pi r^2 \brho(r) \,, 
\label{mNS} 
\\
\bp'(r) &=& -\left(\bp+\brho\right) \frac{m(r)+4 \pi r^3 \bp(r)}{r(r-2m(r))} \,, 
\label{pNS} 
\\
\nu'(r) &=& -\frac{2\bp'(r)}{\bp(r)+\brho(r)} \,, 
\label{nuNS}
\end{eqnarray}
where the prime denotes derivative with respect to $r$. The above system is supplemented by a barotropic EoS in the form $\bp=\bp(\brho)$.

\begin{figure}[htbp]
    \centering
    \includegraphics[width=0.48 \textwidth]{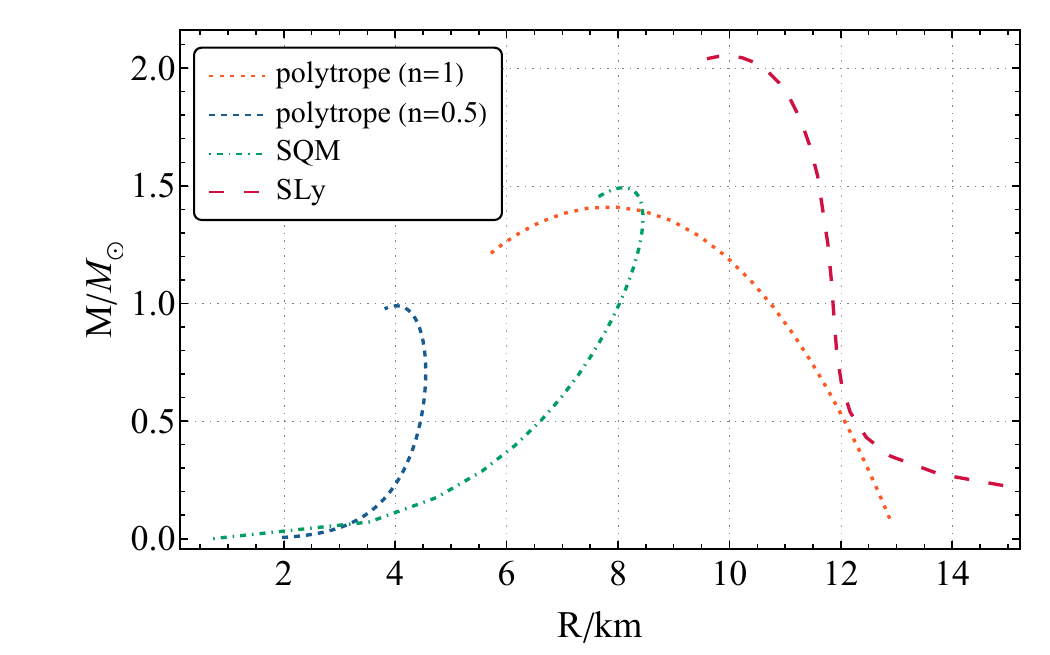}
    \caption{
    NS mass-radius diagram for the EoS used in this work. The EoS for polytropes and SQM depends on a scale factor that can be chosen to fix the mass scale. Dimensionless quantities such as $M/R$ or the dimensionless TLNs are independent of such rescaling.
    }
    \label{fig:mass_radius}
\end{figure}

In what follows we will consider some representative classes of EoS. Our purpose here is not being exhaustive nor realistic in the choice of the EoS catalog, but span different classes to corroborate some EoS-independent properties that will be discussed later on. The first class is a polytropic fluid,
\begin{equation}
\bp(\brho) = \rm{const} \ \brho^{1+\frac{1}{n}}\,,
\end{equation}
where the proportionality constant simply sets a scale, so that dimensionless quantities, like the TLNs, are independent of it. We will consider the polytropic indexes $n=1$ and $n=1/2$, to produce the typical dependence displayed by realistic, nuclear-physics based, EoS. 
We will also consider a representative example of tabulated EoS describing normal $npe\mu$ matter, namely the SLy EoS~\cite{Douchin:2001sv}. 
Finally, we have considered a self-bound strange quark matter~(SQM) EoS as well~\cite{Prakash:1995uw}:
\begin{equation}
\bp(\brho) = \frac{\brho}{3}-4B_{0}\,,
\end{equation}
where the bag constant, $B_0$, again sets a scale but does not alter dimensionless quantities. In this model the density has a jump at the radius of the star, wherein the pressure vanishes. 
For each of these choices of EoS, we obtain the corresponding mass-radius diagram, shown in~\ref{fig:mass_radius}.
Each point on the mass-radius diagram corresponds to different choices for the central density of the star, with its mass and radius are defined by $M=m(R)$ and $\bp(R)=0$, respectively. 

\begin{figure*}[htbp]
    \centering
    \includegraphics[width=0.48 \textwidth]{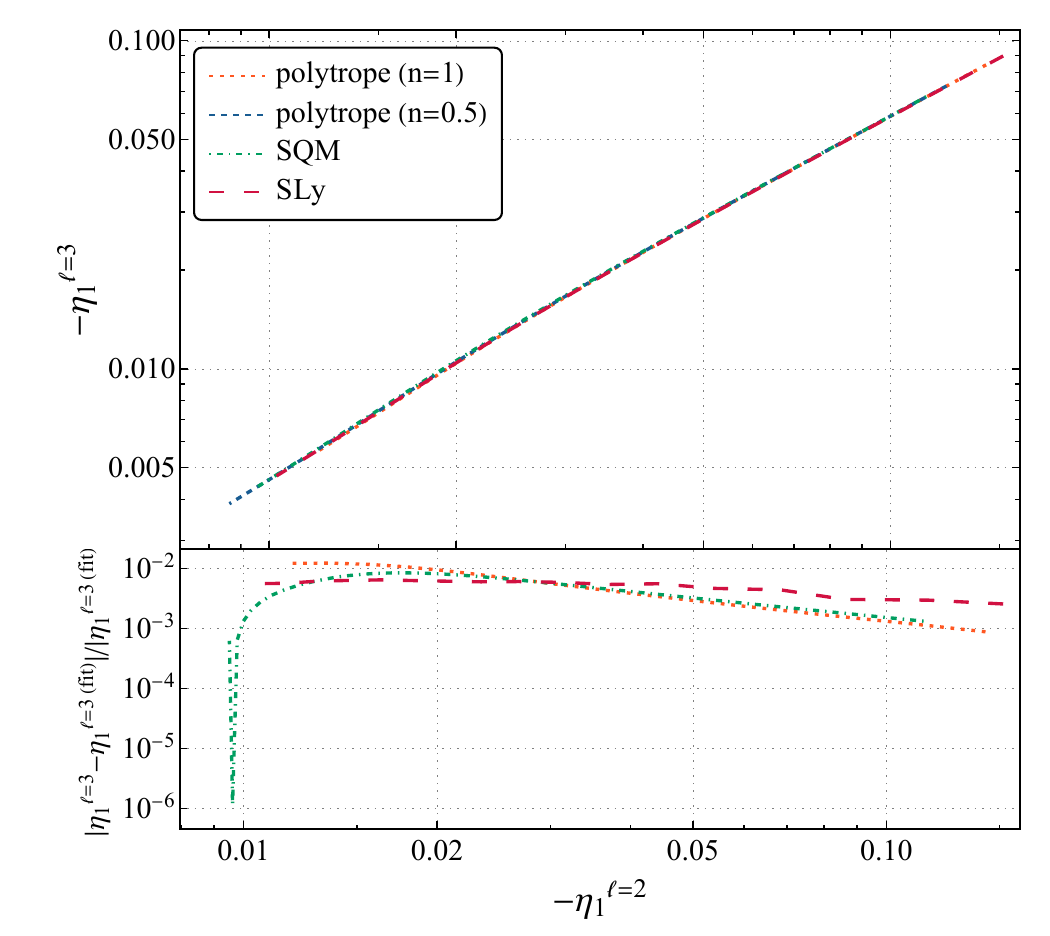}
    \includegraphics[width=0.48 \textwidth]{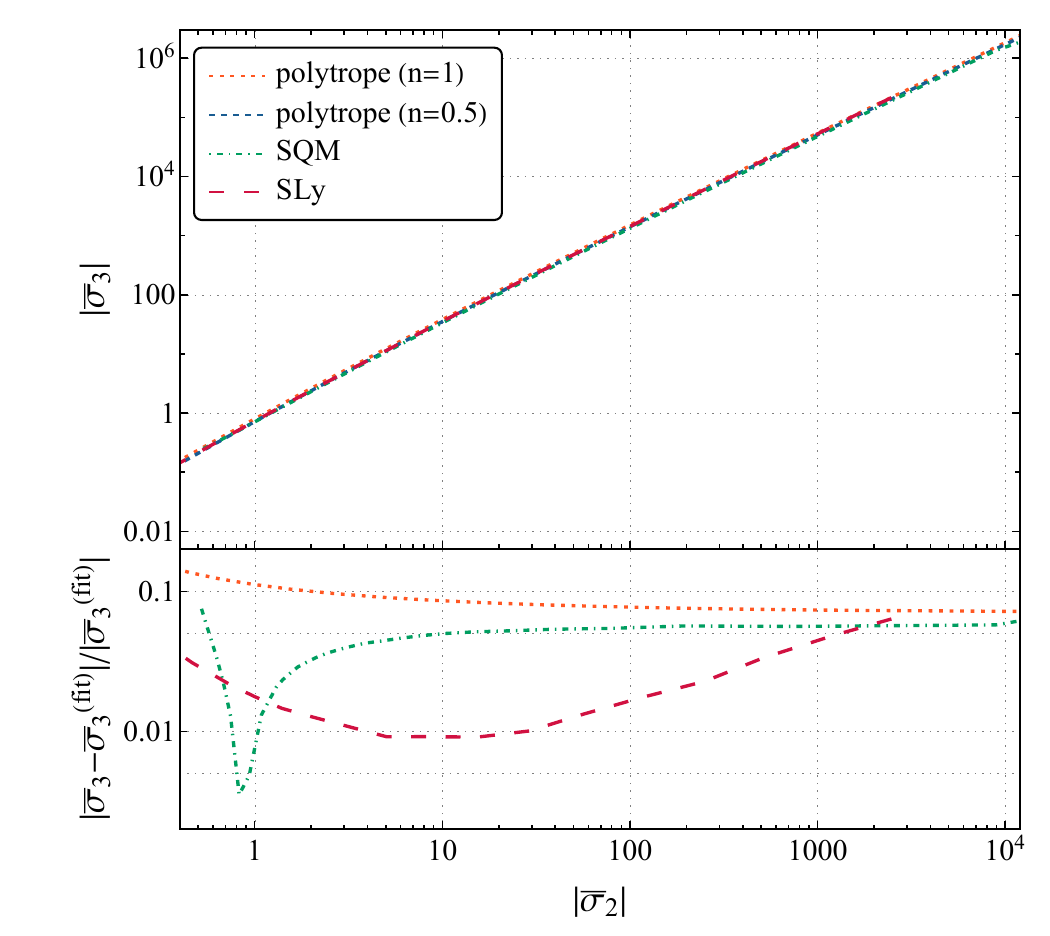}
    \caption{\emph{Left panel:} Quasi-universal relation between the shear viscosities of a NS with different EoS. \emph{Right panel:} Quasi-universal relation between the dimensionless magnetic tidal deformabilities of a NS with different EoS.
    The bottom panel shows the relative difference with respect to the interpolant of the data for the $n=0.5$ polytropic EoS, taken as a reference. Note that, the share viscosity is more universal than the axial TLNs, as the maximum relative difference is $\mathcal{O}(10^{-2})$ for the shear viscosity, while it is one order of magnitude larger for the TLNs.
    }
    \label{fig:universality_magnetic}
\end{figure*}

\subsubsection{Axial sector}

Let us now describe the nonradial perturbations of the NS, again separating the axial and polar sectors. The axial gravitational perturbations can be combined to yield a master function $\bpsi_{\rm RW}$, which satisfies a Regge-Wheeler-like equation inside the star (see, e.g.,~\cite{Kojima:1992ie}):
\begin{align}
e^{\frac{(\nu-\lambda)}{2}}&\dfrac{d}{dr}\left[e^{\frac{(\nu-\lambda)}{2}}\dfrac{d\bpsi_{\rm RW}}{dr}\right] 
+\Bigg[\omega^2-e^{\nu}\Bigg\{\frac{\ell(\ell+1)}{r^2}-\frac{6m(r)}{r^3}
\nonumber 
\\ 
&\qquad \qquad +4\pi\left(\brho-\bp\right)\Bigg\}\Bigg]\bpsi_{\rm RW}=0 \,,
\label{eqNS}
\end{align}
where, in the Regge-Wheeler gauge, the two nonzero axial metric perturbations $h_{0}$ and $h_{1}$ are related to the master function $\bpsi_{\rm RW}$ as, 
\begin{equation}
h_{0}=-\frac{e^{\frac{(\nu-\lambda)}{2}}}{i\omega}\dfrac{d}{dr}(r\bpsi_{\rm RW}) \,;
\quad
h_{1}=e^{\frac{(\lambda-\nu)}{2}}(r\bpsi_{\rm RW})~.
\end{equation}
For each choice of the EoS, we integrate \ref{eqNS} from the center of the NS (where we impose regularity on the master function $\bpsi_{\rm RW}$) to its radius, for different values of the central energy density, in the small-frequency regime. We checked that the results of the integration do not depend on the frequency when $M \omega \ll 1$. 

The solution for the axial master variable, $\bpsi_{\rm RW}(r)$ is substituted into the axial boundary condition for the membrane paradigm, as in \ref{bcax}, to determine the shear viscosity $\eta$ corresponding to the NS within the membrane paradigm. By expanding the shear viscosity in the low-frequency regime, as in \ref{eta}, we find that $\eta_{0}=0$ and $\eta_{1}$ is purely real. This is a comforting consistency check of our framework, since otherwise the NS would have vanishing TLNs, which is definitely incorrect.  
These conditions are also consistent with the fact that a perfect-fluid NS is nondissipative, i.e., totally reflective. Indeed, based on the results of the previous section, the consequence of having $\eta_{0}=0$ for the shear viscosity, corresponds to $|{\cal R}|^{2}=1+\mathcal{O}(\omega^{2}M^{2})$, i.e., perfectly reflecting at low frequencies.

The linear-in-frequency part of the shear viscosity, i.e., $\eta_{1}$, depends on the angular number $\ell$, as well as on the central density and on the choice of the EoS. In the left panel of~\ref{fig:universality_magnetic} we show $\eta_{1}$ for $\ell=3$ versus $\eta_1$ for $\ell=2$ for different EoS. This plot is obtained, for each EoS, by changing the central density. Since $\eta_1$ is dimensionless, for each EoS we obtain a given curve on this diagram. Remarkably, we find that the curves for different EoS lie on the top of each other within very good accuracy, within $1\%$ or better (see bottom panel).  This property is reminiscent of the quasi-universal relations for NSs~\cite{Yagi:2013bca,Yagi:2013awa,Yagi:2016bkt}, in particular those relating the TLNs of different multipolar order~\cite{Yagi:2013sva}.
Indeed, the right panel of~\ref{fig:universality_magnetic} shows the quasi-universal relations between the dimensionless magnetic tidal deformabilities, $\bar{\sigma}_{\ell}$, defined in \ref{sigma2} and \ref{sigma3} for $\ell=2$ and $\ell=3$, respectively, with different EoS. As evident, this plot depicts the expected universality between the magnetic TLNs~\cite{Yagi:2013sva}. 
Interestingly, the approximated universality holds to a greater extent for $\eta_1^{\ell=3}$ vs. $\eta_1^{\ell=2}$ rather than for $\bar\sigma_3$ vs. $\bar\sigma_3$, by roughly one order of magnitude. 

Note that we have derived the magnetic tidal deformability directly through the membrane paradigm, and hence the consistency with the earlier literature is a confirmation that NSs can also be described within our membrane paradigm. Moreover, since our computation is based on the zero-frequency limit of the dynamical perturbation equations, we recover the result for the magnetic TLNs obtained for an \emph{irrotational}~\cite{Landry:2015zfa} rather than for a \emph{static} fluid (see~\cite{Pani:2018inf} for a discussion). 
Although applying different normalizations which involve powers of the compactness can enhance the degree of universality~\cite{Majumder:2015kfa}, the approximate universal relation between $\eta_1^{\ell=3}$ and $\eta_1^{\ell=2}$ is particularly remarkable since the mapping between $\bar\sigma_\ell$ and $\eta_1^{\ell}$ involves nontrivial functions of the compactness (see \ref{k2magnetic} and~\ref{k3magnetic}), so it is not a simple consequence of the quasi-universal relation among $\bar\sigma_3-\bar\sigma_2$. Indeed, as discussed below such universality is absent for the viscosity parameters associated with the polar sector.

\subsubsection{Polar sector}

\begin{figure*}[htbp]
    \centering
    \includegraphics[width=0.48 \textwidth]{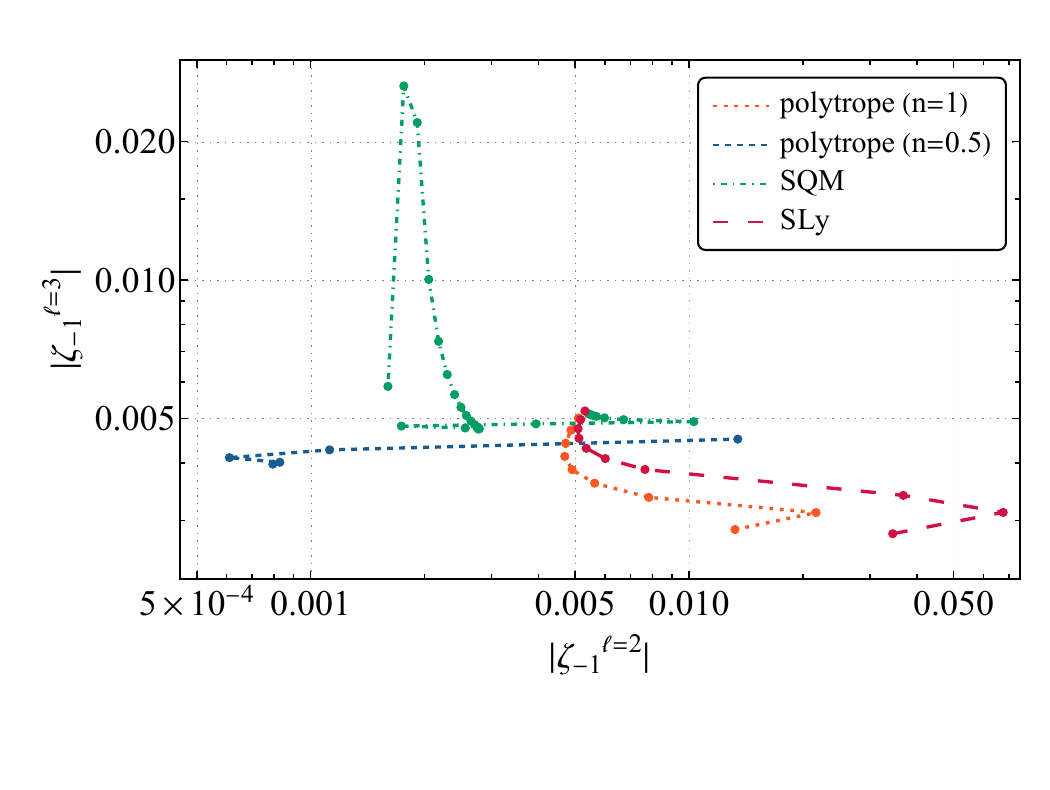}
    \includegraphics[width=0.48 \textwidth]{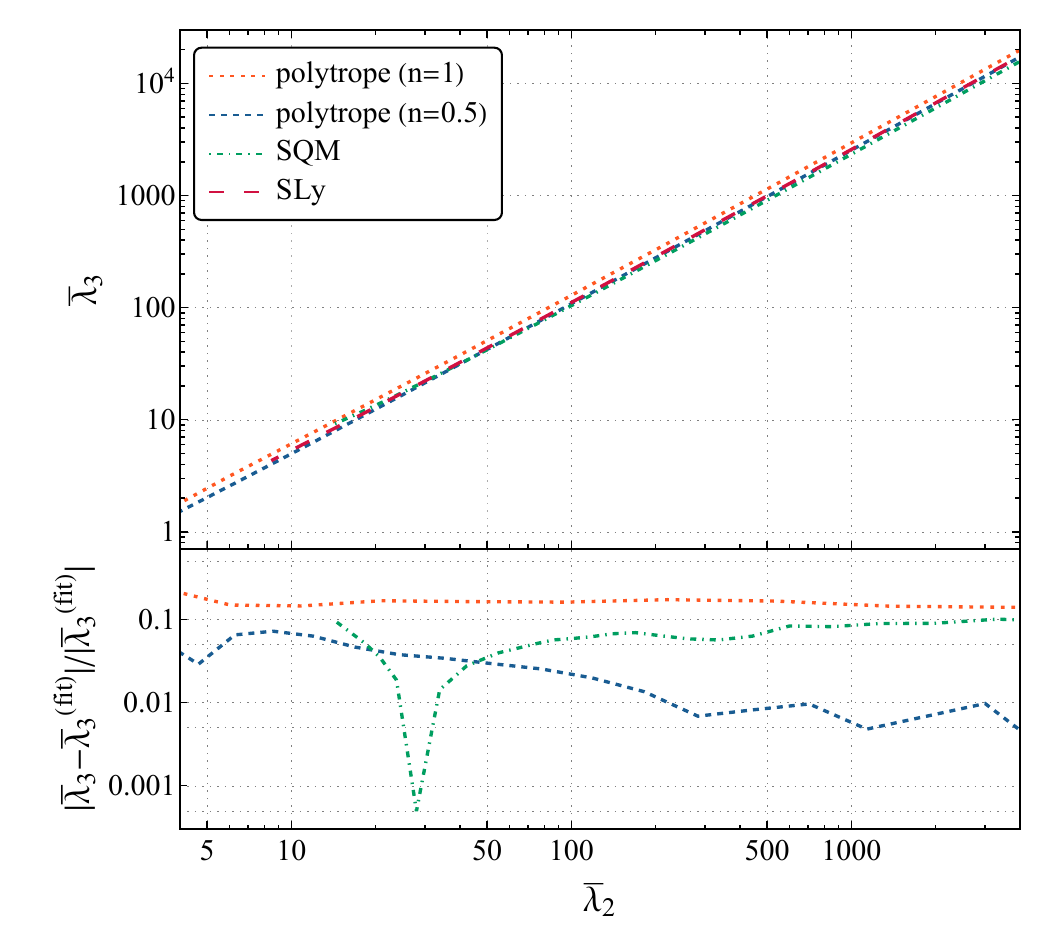}
    \caption{
    \emph{Left panel:} The $\ell=3$ vs. $\ell=2$ bulk viscosities of a NS with different EoS. \emph{Right panel:} Quasi-universal relation between the dimensionless electric tidal deformabilities of a NS with different EoS. The bottom panel shows the relative difference with respect to the interpolant of the data for the SLy EoS, taken as a reference.}
    \label{fig:universality_electric}
\end{figure*}

Turning now to the polar sector, and unlike the axial sector, the nonradial polar perturbations of a NS cannot be described by a single master equation. Rather, it is described by a system of four first order differential equations, two of which are related to fluid perturbations and two are related to metric perturbations. In general, for an isotropic fluid distribution, there are three independent components of polar gravitational perturbations, 
\begin{align}
\delta g_{tt}&=-\int d\omega \sum_{\ell m}r^{\ell}e^{\nu}\hat{H}(r)Y_{\ell m}e^{i\omega t}=e^{\nu-\lambda}\delta g_{rr}~,
\\
\delta g_{tr}&=-\int d\omega \sum_{\ell m}i\omega r^{\ell+1}\hat{H}_{1}(r)Y_{\ell m}e^{i\omega t}=\delta g_{rt}~,
\\
\delta g_{\theta \theta}&=-\int d\omega \sum_{\ell m}r^{\ell+2}\hat{K}(r)Y_{\ell m}e^{i\omega t}=\frac{\delta g_{\phi \phi}}{\sin^{2}\theta}~.
\end{align}
Note that this composition slightly differs from the perturbation variables defined in \ref{polarpert} for the exterior Schwarzschild spacetime (with $e^{\nu}=f$ and $e^{-\lambda}=f$), and is related to them through the following relations: $H=r^{\ell}\hat{H}$, $K=r^{\ell}\hat{K}$ and $H_{1}=-i\omega r^{\ell+1}\hat{H}_{1}$, along with $\omega \to -\omega$. Among these three, we will use $\hat{H}_{1}$ and $\hat{K}$, while $\hat{H}$ can be determined using an algebraic relation between these three. 

The perturbations of the matter sector of a NS depends on the perturbation of the four velocity of the fluid comprising of the NS. These are determined in terms of the metric perturbation $\hat{H}$, along with two additional perturbations $\hat{W}(r)$ and $\hat{V}(r)$, given by, 
\begin{align}
\delta u^{r}&=\int d\omega \sum_{\ell m}i\omega r^{\ell-1}e^{-\frac{\nu+\lambda}{2}}\hat{W}(r)Y_{\ell m}e^{i\omega t}~,
\\
\delta u^{\theta}&=-\int d\omega \sum_{\ell m}i\omega r^{\ell-2}e^{-\frac{\nu}{2}}\hat{V}(r)\partial_{\theta}Y_{\ell m}e^{i\omega t}~,
\\
\delta u^{\phi}&=-\int d\omega \sum_{\ell m}i\omega \frac{r^{\ell-2}}{\sin^{2}\theta}e^{-\frac{\nu}{2}}\hat{V}(r)\partial_{\phi}Y_{\ell m}e^{i\omega t}~.
\end{align}
Instead of the fluid perturbation variable $\hat{V}$, defined above, it is advantageous to introduce another fluid perturbation $\hat{X}$, which is defined as,
\begin{align}
\hat{X}&=\left(\brho+\bp\right)c_{\rm s}^{2}\Big[\frac{1}{r}e^{\frac{\nu-\lambda}{2}}\hat{W}'+\frac{\ell+1}{r^{2}}e^{\frac{\nu-\lambda}{2}}\hat{W}
\nonumber
\\
&+\frac{\ell(\ell+1)}{r^{2}}e^{\frac{\nu}{2}}\hat{V}-e^{\frac{\nu}{2}}\hat{K}-e^{\frac{\nu}{2}}\frac{\hat{H}}{2}\Big]\,.
\end{align}
This relation can be inverted to yield the fluid perturbation variable $\hat{V}$ in terms of the newly defined quantity $\hat{X}$. Thus we have two metric perturbations $\hat{H}_{1}$ and $\hat{K}$, as well as two fluid perturbations $\hat{W}$ and $\hat{X}$. The system of four such first order differential equations reads (see, e.g., \cite{1985ApJ...292...12D,Mondal:2023wwo} for details and for the definitions of the variables):  
%
\begin{widetext}
\begin{eqnarray}
    \hat H_1' &=& \frac{e^\lambda}{r} \hat H + \frac{e^\lambda}{r} \hat K - \frac{(\ell+1) + (2m/r) \ e^\lambda + 4 \pi r^2 e^\lambda (\hat p - \hat \rho)}{r} \hat H_1 - \frac{16 \pi e^\lambda (\hat \rho + \hat p)}{r} \hat V \,, \label{eqH1}\\
    \hat K' &=& \frac{1}{r} \hat H - \left( \frac{\ell+1}{r} - \frac{\nu'}{2} \right) \hat K + \frac{\ell(\ell+1)}{2 r}\hat H_1 - \frac{8 \pi e^{\lambda/2} (\hat p + \hat \rho)}{r} \hat W \,, 
    \label{eqK}\\ 
    \hat W' &=& \frac{e^{\lambda/2}r}{2} \hat H + e^{\lambda/2}r \hat K - \frac{e^{\lambda/2} \ell (\ell+1)}{r} \hat V - \frac{\ell+1}{r} \hat W + \frac{r e^{(\lambda-\nu)/2}}{c_s^2 (\hat \rho + \hat p)} \hat X \,,  
    \label{eqpolar1}\\
    \hat X' &=& - \frac{\ell}{r} \hat X + (\hat p + \hat \rho) e^{\nu/2} \left( \frac{1}{2 r} - \frac{\nu'}{4}\right) \hat H - \frac{e^{\nu/2}\ell(\ell+1)(\hat p + \hat \rho) \nu'}{2 r^2} \hat V + \frac{e^{\nu/2}(\hat p +\hat \rho)}{2} \left[ \frac{\ell(\ell+1)}{2r} +r \omega^2 e^{-\nu} \right] \hat H_1 
    \nonumber \\
    &&~+\frac{e^{\nu/2}(\hat p +\hat \rho)}{2} \left( \frac{3 \nu'}{2} - \frac{1}{r} \right) \hat K - \frac{e^{\nu/2}(\hat p +\hat \rho)}{r} \left[ e^{\lambda/2-\nu} \omega^2 + 4 \pi e^{\lambda/2} (\hat p +\hat \rho) - \frac{r^2}{2} \left(\frac{\nu'}{r^2 e^{\lambda/2}}\right)' \right] \hat W \,, \label{eqpolar4}
\end{eqnarray}
where $c_s^2=(\partial \bp/\partial\brho)$ is the squared speed of sound, and the functions are related by the following algebraic equations
\begin{align}
    \left[ 3 m + \frac{1}{2}(\ell-1)(\ell+2) r + 4 \pi r^3 \hat p \right] \hat H &= 8 \pi e^{-\nu/2} r^3 \hat X - \left[ - e^{-\nu-\lambda} r^3 \omega^2 +\frac{1}{2} \ell (\ell+1) \left(m+4 \pi r^3 \hat p\right)\right] \hat H_1\nonumber \\
    &+ \Bigg[\frac{(\ell-1)(\ell+2)}{2}  r -e^{-\nu} r^3 \omega^2 - \frac{e^\lambda}{r} \left(m+4 \pi r^3 \hat p\right) \left(3m-r+4 \pi r^3 \hat p\right) \Bigg] \hat K \,, \\
    \omega^2 (\hat p+\hat \rho) \hat V &= e^{\nu/2} \hat{X} - \frac{e^\nu (\hat p+\hat \rho)}{2} \hat H + \frac{e^{\nu-\lambda/2} \hat p'}{r} \hat W \,. 
\end{align}
\end{widetext}
For each choices of the EoSs, we integrate \ref{eqH1}-\ref{eqpolar4} from the center of the NS to the radius of the NS, with different values of the central energy density. At the center we impose regularity of the perturbations, namely $\hat{Y}(r) = \sum_i \hat{Y}(i)\,r^i$, where $\hat{Y}=(\hat H_1,\hat K,\hat W,\hat X)$ collectively denotes all the perturbation variables.
The coefficients $\hat{Y}_{i}$ are fixed by solving the above linear perturbation equations order by order, near the center, e.g., to first order we obtain the following two relations:
\begin{align}
    &\hat H_1(0)=\frac{2 \ell \hat K(0) +16 \pi (\hat \rho_0 + \hat p_0) \hat W(0)}{\ell (\ell+1)} \,, \\
    &\hat X(0)=(\hat \rho_0 + \hat p_0) e^{\nu_0/2}   \nonumber \\
    &\times \Bigg[ \left( \frac{4 \pi}{3} (\hat \rho_0 + 3 \hat p_0) - \frac{\omega^2}{\ell} e^{-\nu_0}\right)\hat W(0)+ \frac{1}{2} \hat K(0) \Bigg] \,.
\end{align}
We set the values of $\hat W(0)$ and $\hat K(0)$ such that, upon numerical integration, we get $\hat X(R)=0$ \cite{1985ApJ...292...12D,Mondal:2023wwo}. For this purpose, we perform two independent integrations with $\hat W(0)=1, \hat X(0)= (\hat \rho_0+\hat p_0)$ and $\hat W(0)=1, \hat X(0)= -(\hat \rho_0+\hat p_0)$, respectively, and obtain two values of $\hat X(R)$ at the NS radius, namely $\hat X_+(R)$ and $\hat X_-(R)$. We linearly combine the two solutions as $\hat X(R) = \hat X_+(R) + c \ \hat X_-(R)$ and determine the value of the coefficient $c$ to ensure that $\hat X(R)=0$.

The solutions for $\hat H_1$ and $\hat K$ are then converted into a solution for the polar master variable, $\bpsi_{\rm Z}(r)$, and substituted into the polar boundary condition for the membrane paradigm, as in \ref{bcpol}, to determine the bulk viscosity $\zeta$ of the fictitious membrane that matches the NS interior. 
By expanding the bulk viscosity in the low-frequency regime, as in \ref{zeta}, we find that $\zeta_{-1}$ is purely real and $\zeta_{0}=0$. 
Again, these viscosity parameters correspond to $|{\cal R}|^2=1$ up to quadratic order in $\omega$. The left panel of~\ref{fig:universality_electric} shows the bulk viscosity for $\ell=2$ and $\ell=3$ for different EoS. Differently from the shear viscosity shown in~\ref{fig:universality_magnetic}, the bulk viscosity is not universal but depends on the NS EoS.
This should not come as a surprise given the nontrivial relation between $\zeta_{-1}$ and the TLNs, both being complicated functions of the compactness of the NS.
Indeed, as a confirmation of our framework we can derive the standard electric TLNs of the NS directly from the membrane paradigm, by using~\ref{k2E} and~\ref{k3E} where the mass, the radius and the bulk viscosity are functions of the central density and the EoS.
The right panel of~\ref{fig:universality_electric} shows the quasi-universal relations between the dimensionless electric tidal deformabilities for $\ell=2$ and $\ell=3$ with different EoS. These results agree with the electric-type Love numbers in~\cite{Binnington:2009bb} for polytropic EoS and with the results of~\cite{Yagi:2013awa} for what concerns the approximate universality. 
We do not exclude some other possible nontrivial quasi-universal relations involving the bulk viscosity $\zeta_{-1}$, the angular number $\ell$, and powers of the compactness.

To summarize, our membrane paradigm correctly reproduces the TLNs of NSs and has unveiled a novel, nontrivial, quasi-universal relation between the shear viscosity parameters of the membrane in the axial sector. This leads to a quasi-universal relation between the axial TLNs, even though the axial TLNs are related to the shear viscosity parameters through complicated functions of the compactness. On the other hand, the quasi-universality between the bulk viscosity parameters is broken, but still the polar TLNs derived from the bulk viscosity depicts quasi-universality. 

\subsubsection{Comment on the boundary conditions in case of discontinuities}

When matching the perturbations between the NS interior and the exterior Schwarzschild spacetime, one should be careful with potential discontinuities of the energy density across the radius of the NS. Indeed, while the pressure must be continuous and $\hat p_0(R)=0$, the energy density might have a discontinuity, as it happens for the SQM EoS. 
Indeed, in that case $\hat\rho_0=F(r)\Theta(R-r)$, where $F$ is some function and $\Theta$ is the Heaviside step function. Therefore, $\partial_r\hat\rho_0\sim \delta(r-R)$, which might imply discontinuities in the perturbation variables. This is indeed the case when computing \emph{static} electric TLNs~\cite{Damour:2009vw} for incompressible fluids or for SQM~\cite{Hinderer:2009ca}.
Interestingly, this does not occur here when solving the \emph{dynamical} linearized equations. Indeed, the boundary conditions depend on $\hat H_1$ and $\hat K$ which are governed by \ref{eqH1} and \ref{eqK}. Since these first-order equations do not depend on the speed of sound (and hence on $\partial_r\hat\rho_0$), no discontinuity appears in the gravitational perturbation upon integration. On the other hand, the fluid perturbation variable $\hat W$ is discontinuous in case $\hat\rho_0$ has a jump, as it depends on the sound speed, and hence on $\partial_r\hat\rho_0$. The other fluid perturbation variable $X$ has no jump at the surface of the NS. 
In general, no such discontinuity appears in the axial sector as well, since the axial perturbation equations do not explicitly depend on the background energy density, and neither on the sound speed.

\subsection{TLNs of gravastars from membrane paradigm}

In this section, we derive the mapping between the membrane paradigm and a concrete model of ECO. Due to its simplicity and possibility to be solved fully analytically, we focus on the gravastar model. The latter is described by a de Sitter interior matched to the Schwarzschild exterior with some matter distribution at the interface~\cite{Chapline:2000en,Mazur:2004fk}. The stability of such a model under radial \cite{Visser:2003ge} and non-radial perturbations in both the axial and polar sectors~\cite{Chirenti:2007mk,Pani:2009ss} have been studied extensively.  
The thin-shell gravastar~\cite{Visser:2003ge} is described by the following line element 
\begin{align}
ds^{2}&=-f(r)dt^{2}+\frac{dr^{2}}{g(r)}+r^{2}d\Omega^{2}~,
\\
f(r)&=h(r)=\left(1-\frac{2M}{r}\right)\qquad ~ \qquad r>R
\nonumber
\\
&=\alpha h(r)=\alpha \left(1-\frac{2 M_{\rm v}}{R^3}r^{2}\right) \qquad r<R
\end{align}
where $M=M_{\rm v}+(M_{\rm s}^{2}/2R)+M_{\rm s}\sqrt{1-(2M_{\rm v}/R)}$ is the total mass, with $M_{\rm s}=4\pi R^{2}\Sigma$. Indeed, at $r=R=2M(1+\epsilon)$ there is a discontinuity due to a thin shell with nonzero surface energy density $\Sigma$. This is a physical quantity not to be confused with the surface energy density of the fictitious membrane in our membrane paradigm. The parameter $\alpha$ appearing in the $g_{tt}$ component of the metric is determined in terms of the total mass $M$ and the volume mass $M_{\rm v}$ of the gravastar, such that, $\alpha=\{1-(2M/R)\}\{1-(2M_{\rm v}/R)\}^{-1}$, which ensures continuity of the metric across the surface of the gravastar. 

\begin{figure}[t]
    \centering
    \includegraphics[width=0.48 \textwidth]{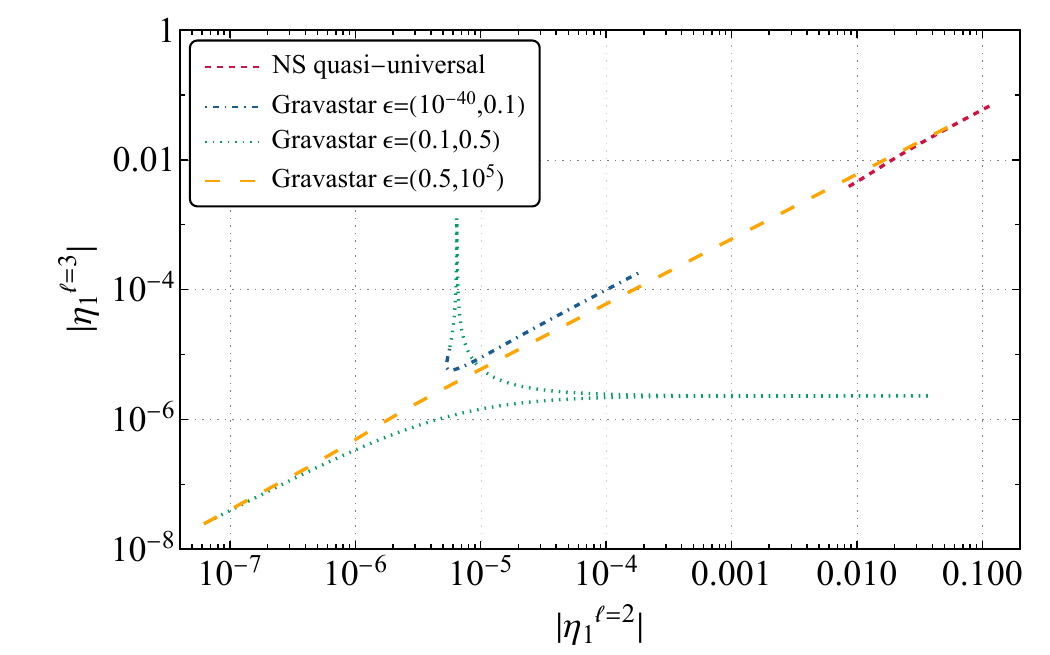}
    \caption{Shear viscosities of a gravastar with different compactness, where $\epsilon = (10^{-40},10^5)$, compared to the quasi-universal relation of a NS with different EoS.
    For clarity, we show the gravastar curve in difference ranges of compactness with different styles and colors.
    }
\label{fig:gravastar}
\end{figure}

Given this spacetime, one can determine the master equations satisfied by the axial and the polar gravitational perturbations in the interior and the exterior of the gravastar~\cite{Pani:2009ss}. Intriguingly, both the axial and the polar perturbations in the interior of the gravastar are determined by a single master equation~\cite{Pani:2009ss} 
\begin{align}
\dfrac{d^{2}\Psi_{\rm int}}{dr_{*}^{2}}+\left(\omega^{2}-V_{\rm int}\right)\Psi_{\rm int}=0~, 
\end{align}
where, $V_{\rm int}=\{\ell(\ell+1)/r^{2}\}f(r)$ and $\Psi_{\rm int}$ is a master radial function (related to either axial or polar perturbations by expressions given in~\cite{Pani:2009ss}). The above equation can be solved exactly. The solution which is regular at $r=0$ reads~\cite{Pani:2009ss}
\begin{align}
\Psi_{\rm int}&=r^{\ell+1}{f(r)^{i\frac{M\omega}{\sqrt{C}}}}
\nonumber 
\\ 
&\times\,_{2}F_{1}\left(\frac{\ell+2+\frac{2iM\omega}{\sqrt{C}}}{2},\frac{\ell+1+\frac{2iM\omega}{\sqrt{C}}}{2},\ell+\frac{3}{2},\frac{Cr^2}{4M^2}\right)\,, 
\nonumber 
\end{align}
%
where $C=(2M/R)^{3}=(1+\epsilon)^{-3}$. Given this interior perturbation variable, which is the proxy for the Regge-Wheeler variable outside the NS, one can use the boundary condition for the axial sector, as presented in \ref{bcax}, and substitute the expression for $\Psi_{\rm int}$ in order to determine the following expression for the linear-in-frequency shear viscosity of the fictitious membrane fluid mimicking the physics of a gravastar to the exterior spacetime,
\begin{widetext}
\begin{equation}
    \eta=-\frac{iM\omega(1+\epsilon)^{3}}{8\pi(2+\ell)\epsilon 
    \left(\frac{(1+\epsilon)\left\{-2+\ell+\epsilon\left(\ell+1\right)\right\}}{2\epsilon-1}+\frac{(1+\ell)}{2} \frac{~_2F_1^{\rm R}\left(\frac{3+\ell}{2}, \frac{4+\ell}{2}, \frac{5}{2}+\ell, \frac{1}{1+\epsilon}\right)}{~_2F_1^{\rm R}\left(\frac{1+\ell}{2}, \frac{2+\ell}{2}, \frac{3}{2}+\ell, \frac{1}{1+\epsilon}\right)} \right)} + \mathcal{O}(M^2\omega^2)
    \simeq \frac{\log(\epsilon)}{16\pi}i\omega M~,
    \label{etagravastar}
\end{equation}
\end{widetext}
where in the last step we assumed $\epsilon\to0$. In the above expression, $_{2}F_1^{\rm R}(a,b;c;z)$ is the regularized hypergeometric function, related to the hypergeometric function $_{2}F_1(a,b;c;z)$ by the division of $\Gamma(c)$. It is evident from \ref{etagravastar} that the shear viscosity $\eta$ for a gravastar is also proportional to $\omega$, which immediately shows that $\eta_{0}=0$ in the expansion of the shear viscosity, as in \ref{eta}. This is another consistency check of the framework presented here, since a gravastar is a non-dissipative, but reflective system with non-zero TLNs. 
In \ref{fig:gravastar} we show the shear viscosity of a gravastar for different compactnesses where $\epsilon$ spans from $10^{-40}$ to $10^5$. The shear viscosity shows some divergencies, which depend on the angular number of the perturbation $\ell$. Curiously, in the limit of small compactness (corresponding to large values of $\epsilon$), the gravastar's shear viscosity recovers the quasi-universal relation for the NSs.

Furthermore, by plugging the above expression for $\eta_1$ from \ref{etagravastar} in \ref{tlnax}, we derive the axial TLN associated with the $\ell=2$ mode of a gravastar,
\begin{equation}\label{tlnaxial}
k_{2}^{\rm M} = \frac{32}{5 \left(43-12 \log2+18 \log \epsilon\right)} + \mathcal{O}(\epsilon) \,,
\end{equation}
which coincides with the previous results regarding the axial TLNs of gravastars~\cite{Cardoso:2017cfl}. This is another nontrivial consistency check of our formalism.
Interestingly, the leading order behavior of $k_{2}^{\rm M}\sim \{32/90\log(\epsilon)\}$ can also be obtained from \ref{eta1} and \ref{tlnax}. First of all, from \ref{eta1} and \ref{etagravastar} it follows that the linear-in-frequency reflectivity associated with the axial sector of a gravastar is given by $\mathcal{R}_{1}^{\rm ax}\simeq -2\log(\epsilon)$. Substituting this into \ref{tlnax} yields the leading order behavior of magnetic TLNs to be $k_{2}^{\rm M}\sim \{1/90\log(\epsilon)\}$, which coincides with \ref{tlnaxial}, modulo an overall normalization factor.

A similar strategy can be employed also for the polar perturbations. Here we use the same interior solution for the gravastar, but now substitute it in the relevant junction condition associated with the polar perturbation of the membrane, namely \ref{bcpol}. In addition, by inserting the expressions for the shear viscosity, we obtain the bulk viscosity of a gravastar when mapped to the membrane paradigm as,
\begin{widetext}
\begin{equation}\label{zetagravastar}
\zeta = \frac{i (1+\frac{\gamma_{\ell}}{2})}{64 M \omega \pi \epsilon (1+\epsilon)[3+\gamma_{\ell}(1+\epsilon)]^2}
\frac{A~_2F_1^{\rm R} \left(\frac{1+\ell}{2}, \frac{2+\ell}{2}, \frac{3}{2}+\ell, \frac{1}{1+\epsilon}\right)+B~_2F_1^{\rm R} \left(\frac{3+\ell}{2}, \frac{4+\ell}{2}, \frac{5}{2}+\ell, \frac{1}{1+\epsilon}\right)}{C~_2F_1^{\rm R} \left(\frac{1+\ell}{2}, \frac{2+\ell}{2}, \frac{3}{2}+\ell, \frac{1}{1+\epsilon}\right)+D~_2F_1^{\rm R} \left(\frac{3+\ell}{2}, \frac{4+\ell}{2}, \frac{5}{2}+\ell, \frac{1}{1+\epsilon}\right)} + \mathcal{O}(M \omega)^0 \,,
\end{equation}
where, the quantities $A$, $B$, $C$ and $D$ are given in terms of the compactness by 
\begin{eqnarray}
A &=&-2(1 + \epsilon) \left\{2\gamma_{\ell} (1 + \epsilon) \left[3 + \gamma_{\ell} + 3 (2 + \gamma_{\ell}) \epsilon^2 + 4\gamma_{\ell} \epsilon^3\right] 
+\ell\left[3 + \gamma_{\ell} (1 + \epsilon)\right] \left[3 + 12 \epsilon + \gamma_{\ell} (1 + \epsilon) \left(1 + 2 \epsilon + 4 \epsilon^2\right)\right]\right\}\,, 
\nonumber
\\
B &=& -(1 +\ell) (2 + \ell) \left[3 + \gamma_{\ell} (1 + \epsilon)\right] \left[3 + 12 \epsilon + \gamma_{\ell} (1 + \epsilon) \left(1 + 2 \epsilon + 4 \epsilon^2\right)\right]  \,, 
\nonumber 
\\
C &=& 2 (1 + \epsilon) \left\{\gamma_{\ell}(1 + \epsilon) +\ell \left[3 + \gamma_{\ell} (1 + \epsilon)\right]\right\} \,, \nonumber 
\\
D &=& (1 + \ell) (2 + \ell) \left[3 + \gamma_{\ell} (1 + \epsilon))\right] \,,\nonumber
\end{eqnarray}
\end{widetext}
and $\gamma_{\ell}$ was defined earlier, see the discussion around \ref{genBCaxial}. Also in this case the bulk viscosity for gravastar scales as $1/(\omega\epsilon)$, as expected from our previous discussion. Thus, also for gravastars, we have nonzero $\zeta_{-1}$, with $\zeta_{0}=0$, whose small-$\epsilon$ behavior is given by:
\begin{align}
\zeta_{-1}^{\ell=2}&=- \frac{3 i}{448 \pi \epsilon} - \frac{3 i [3 + \log 16 - 2 \log \epsilon]}{3136 \pi}~,\label{gravzetam2} \\
\zeta_{-1}^{\ell=3}&=- \frac{3 i}{416 \pi \epsilon} - \frac{i [-28 + 30 \log 2 - 15 \log \epsilon]}{5408 \pi}~,
\end{align}
for the $\ell=2$ and $\ell=3$ modes, respectively. Substituting \ref{zetagravastar} into the expression for the electric TLN, as in \ref{kER}, we obtain the polar TLN for the $\ell=2$ mode of a gravastar,
\begin{equation}
k_{2}^{\rm E} = \frac{16}{5 \left(23-6 \log2+9 \log \epsilon\right)} + \mathcal{O}(\epsilon) \,,
\end{equation}
which again coincides with the result of Ref.~\cite{Cardoso:2017cfl}. Like in the axial sector, it follows from the comparison of \ref{gravzetam2} and \ref{zetam1l2}, that at leading order, $\mathcal{R}_{1}^{\rm polar}\simeq-2\log(\epsilon)$, and hence substitution of the same in \ref{kER} yields the leading order behavior of the polar TLN associated with $\ell=2$ mode, as in the above expression. This gives a further nontrivial consistency check of our results. Also in the case of gravastars both the shear and the bulk viscosities associated with the membrane paradigm are purely imaginary, consistently with what generically expected for a nondissipative object.

\section{Conclusions} \label{sec:conclusions}
In conclusion, we have presented a model-agnostic framework for studying the tidal deformability of compact objects using the membrane paradigm. This approach encodes the tidal properties of compact objects in terms of viscosity parameters associated with a fictitious membrane. Besides BHs, our framework  accommodates any material compact object, including NSs and ECOs, provided the bulk and shear viscosity parameters of the membrane are nontrivial frequency-dependent functions.

We derived general expressions for the electric and magnetic TLNs in terms of these viscosity parameters and analyzed their behavior with respect to the compactness of the object. This analysis also allowed us to identify the conditions under which a logarithmic dependence, previously observed in certain models, emerges. Likewise, we showed on general grounds that the axial and polar TLNs are the same provided the reflectivity of the object for axial and polar perturbation is the same to linear order in the frequency.

An important outcome of our study is a novel quasi-universal relation among the membrane shear viscosity for different $\ell$'s associated to NSs, which are almost insensitive to the EoS.  This is reminiscent of (and in fact more accurate than) a similar quasi-universal relation among the NS magnetic TLNs with different $\ell$'s, but it is a nontrivial feature, given the complicate mapping between these two quantities. Indeed, we did not find a similar universality for the bulk viscosity, despite of the fact that the electric TLNs with different $\ell$'s enjoy quasi-universal relations.
As for other quasi-universal relations~\cite{Yagi:2016bkt}, it would be very interesting to investigate the underlying reason for this approximate universality.

Overall, our framework offers a robust and flexible tool for investigating the tidal properties of compact objects.
Natural extensions are adding rotation (perturbatively in the angular momentum~\cite{Saketh:2024ojw} or generically), and including possible beyond-GR effects.
It might also be interesting to investigate whether this new framework can shade light on dynamical tidal effects, which are currently under intense scrutiny.
Finally, one could incorporate our results in parameter-estimation pipelines, to constrain the viscosity parameters of the membrane and develop model-agnostic tests of NSs and ECOs.

\begin{acknowledgments}
E.M. is supported by the European Union’s Horizon Europe research and innovation programme under the Marie Skłodowska-Curie grant agreement No. 101107586.
E.M. acknowledges funding from the Deutsche Forschungsgemeinschaft (DFG) - project number: 386119226.
S.C. acknowledges the warm hospitality at the Albert-Einstein Institute, where a part of this work was performed, which was supported by the Max-Planck-India mobility grant. Research of S.C. is supported by the MATRICS and the Core research grants from SERB, Government of India (Reg. Nos. MTR/2023/000049 and CRG/2023/000934). S.C. also thanks ``Beyond the horizon" school at ICTS, where a part of this work was done.
P.P. is partially supported by the MUR FIS2 Advanced Grant ET-NOW (CUP:~B53C25001080001), the MUR PRIN Grant 2020KR4KN2 ``String
Theory as a bridge between Gauge Theories and Quantum Gravity'', by the MUR FARE programme (GW-NEXT, CUP:~B84I20000100001), and by the INFN TEONGRAV initiative. 
\end{acknowledgments}

\appendix
\labelformat{section}{Appendix #1} 
\labelformat{subsection}{Appendix #1}
\begin{widetext}

\section{Boundary conditions for gravitational perturbations of the membrane}\label{appbcaxial}

\subsection{Axial sector}

Here we present the derivation of the boundary condition for the perturbation variable $h_{0}$ arising out of the junction conditions associated with the membrane. From the nonzero components of \ref{pertaxial}, we obtain two equations, one is the expression for the perturbed velocity $\delta u^{\phi}$, presented in \ref{velpertaxial}, and the other one is a relation between $h_{0}$ and $h_{1}$,
\begin{align}
h_{1}(R)&= \frac{-2\eta}{\left(\rho_{0}+p_{0}\right)f(R)} \Bigg[\frac{1}{2}\sqrt{f(R)}\left[i\omega h_{1}(R)+h_{0}'(R)\right]
-\left(\frac{f'(R)}{2\sqrt{f(R)}}+\frac{2\sqrt{f(R)}}{R}\right)h_{0}(R) - 8\pi \rho_{0}(R)h_{0}(R)\Bigg]~.
\end{align}
Note that, using \ref{rho0} and \ref{p0} in the above equation, the latter can be reduced to the following form,
\begin{align}
h_1(R)=\frac{16 \pi  R \eta \left[f(R) h_{0}'(R)-h_{0}(R) f'(R)\right] }{f(R) \left[-R f'(R)+2 f(R)-16i\pi  R \omega \right]}\,.
\end{align}
In order to eliminate $h_{1}$, and express the above boundary condition in terms of $h_{0}$ alone, we use the following relation between $h_{1}$ and $h_{0}$, obtained from perturbation of the vacuum Einstein's equations, 
\begin{align}
h_{1}=-\frac{i\omega r^{4}}{(\ell+2)(\ell-1)f(r)}\frac{d}{dr}\left(\frac{h_{0}}{r^{2}}\right)+\mathcal{O}(M^{2}\omega^{2})~.
\end{align}
Therefore, we obtain the following boundary condition for the axial metric perturbation $h_{0}$, which reads,
\begin{align}
\frac{i\omega R^{4}}{(\ell+2)(\ell-1)f(R)}\frac{d}{dr}\left(\frac{h_{0}}{r^{2}}\right)_{R}&=\frac{2\eta}{\left(\rho_{0}+p_{0}\right)f(R)} 
\Bigg[\frac{1}{2}\sqrt{f(R)}\left[-i\omega \times \frac{i\omega R^{4}}{(\ell+2)(\ell-1)f(R)}\frac{d}{dr}\left(\frac{h_{0}}{r^{2}}\right)_{R}+h_{0}'(R)\right]
\nonumber
\\
&-\left(\frac{f'(R)}{2\sqrt{f(R)}}+\frac{2\sqrt{f(R)}}{R}\right)h_{0}(R) - 8\pi \rho_{0}(R)h_{0}(R)\Bigg]~.
\end{align}
Ignoring contributions coming from terms $\mathcal{O}(M^{2}\omega^{2})$, the above boundary condition can be re-expressed as,
\begin{align}
\frac{i\omega R^{4}}{(\ell+2)(\ell-1)f(R)}\frac{d}{dr}\left(\frac{h_{0}}{r^{2}}\right)_{R}&=\frac{2\eta}{\left(\rho_{0}+p_{0}\right)f(R)} 
\Bigg[\frac{1}{2}\sqrt{f(R)}h_{0}'(R)
\nonumber
\\
&-\left(\frac{f'(R)}{2\sqrt{f(R)}}+\frac{2\sqrt{f(R)}}{R}\right)h_{0}(R) - 8\pi \rho_{0}(R)h_{0}(R)\Bigg]~,
\end{align}
which also has the following equivalent and simplified form,
\begin{align}
\left(\frac{h'_{0}}{h_{0}}\right)_{R}=\frac{1}{16 f(R)^2}\left[16 f(R) f'(R)+\frac{i \omega  \left(R f'(R)-2 f(R)\right)^2}{\pi  \eta  \left(l^2+l-2\right)}\right]~.
\end{align}
This is the result we have used in order to arrive at \ref{genBCaxial} in the main text. Subsequent simplification yields the boundary condition given in \ref{BCaxialx}, for the membrane in the axial sector. 


\subsection{Polar sector}
In this section, we detail the derivation of the polar boundary condition presented in \ref{BCH}. We first report the full expressions for $\rho_1$ and $p_1$, the first-order expansions in $\omega$ of which are given in. \ref{eq:rho1} and \ref{eq:p1}:
\begin{equation}\label{rho1full}
\rho_1=f(R)( 3 \cos 2 \theta+1)\frac{aH(R)+bH'(R)}{c}\;,
\end{equation}
where
\begin{align}
a &=(3 M-R) \Big[-16 \pi  \eta  R^4 \omega ^3 \left(30 M^2-31 M R+10 R^2\right)+128 \pi  \eta  \omega  (2 M-R) \left(3 M^3-3 M^2 R+5 M R^2-3 R^3\right)-
\nonumber
\\
&\quad~4 i R^2 \omega ^2 \left(6 M^3+28 M^2 R-26 M R^2+5 R^3\right)+i R^6 \omega ^4 (3 M-R)-24 i (4 M-R) (R-2 M)^2-16 \pi  \eta  R^8 \omega ^5\Big]\,, 
\nonumber
\\
b &=4 R \left(6 M^2-5 M R+R^2\right) \Big[16 \pi  \eta  \omega  (2 M-R) \left(3 M^2+6 M R-2 R^2\right)-i R^2 \omega ^2 \left(3 M^2-7 M R+3 R^2\right)+\nonumber
\\
&\quad~16 \pi  \eta  R^4 \omega ^3 (3 M-R)+6 i (R-2 M)^2\Big] \,, 
\nonumber 
\\
c &= 64 \pi  R \left(3 \cos ^2\theta-1\right) (2 M-R) \left(6 M+R^3 \omega ^2-3 R\right) \left(4 M+R^3 \omega ^2-2 R\right) [R (16 \pi  \eta  R \omega +i)-3 i M]\,,
\end{align}
and
\begin{equation}\label{p1full}
p_1= \frac{d H(R)-gH'(R)}{h}\;,
\end{equation}
where
\begin{align}
d &=-24 \left(6 M^2-5 M R+R^2\right)^2+R^6 \omega ^4 \left[256 \pi ^2 \zeta  \eta  \left(30 M^2-31 M R+10 R^2\right)-93 M^2+58 M R-9 R^2\right]+\nonumber
\\
&\quad~16 i \pi  R^4 \omega ^3 \left[3 M^3 (8 \zeta +19 \eta )+2 M^2 R (56 \zeta -39 \eta )+M R^2 (47 \eta -104 \zeta )+10 R^3 (2 \zeta -\eta )\right]-4 R^2 \omega ^2 \Big[-18 M^4+\nonumber
\\
&\quad~132 M^3 R-129 M^2 R^2+512 \pi ^2 \zeta  \eta  (2 M-R) \left(3 M^3-3 M^2 R+5 M R^2-3 R^3\right)+44 M R^3-5 R^4\Big]-\nonumber
\\
&\quad~128 i \pi  \omega  \left[\eta  \left(9 M^5-9 M^4 R+18 M^3 R^2-30 M^2 R^3+17 M R^4-3 R^5\right)-3 \zeta  R^2 (4 M-R) (R-2 M)^2\right]-\nonumber
\\
&\quad~16 i \pi  R^8 \omega ^5 (\zeta +4 \eta ) (3 M-R)+256 \pi ^2 \zeta  \eta  R^{10} \omega ^6\,, 
\nonumber
\\
g &= 4 i R \omega  (2 M-R) \Big\{16 \pi  \left[\eta  \left(3 M^2+3 M R-2 R^2\right)^2+6 \zeta  R^2 (R-2 M)^2\right]-16 \pi  R^4 \omega ^2 \Big[3 M^2 (\zeta -\eta )-M R (7 \zeta +3 \eta )+\nonumber
\\
&\quad~R^2 (3 \zeta +2 \eta ) \Big]-i R^2 \omega  \left[-9 M^3+256 \pi ^2 \zeta  \eta  (2 M-R) \left(3 M^2+6 M R-2 R^2\right)-6 M^2 R+9 M R^2-2 R^3\right]-\nonumber
\\
&\quad~i R^6 \omega ^3 \left(256 \pi ^2 \zeta  \eta -1\right) (3 M-R)\Big\}\,, 
\nonumber 
\\
h &=-i 64 \pi  R^2 f(R) \left(6 M+R^3 \omega ^2-3 R\right) \left(4 M+R^3 \omega ^2-2 R\right) [R (16 \pi  \eta  R \omega +i)-3 i M] \,.
\end{align}
The displacement $R_1$, defined in \ref{eq:R1} depends on $H_1$. In the main text, \ref{H1} provides the first-order expansion in $\omega$ of $H_1$ as a function of  $H\;,H’$ and $l$. Here, we employ the full expression for $H_1$ at $r=R$ and $l=2$, retaining all orders in $\omega$
\begin{equation}
    H_1(R)=\frac{i \omega  \Big\{ \left[R^4 \omega ^2 (19 M-5 R)-8 M \left(3 M^2-6 M R+2 R^2\right)\right]H(R)-4 R (2 M-R)  \left(3 M^2+3 M R+R^4 \omega ^2-2 R^2\right)H'(R)\Big\}}{4 \left(6 M+R^3 \omega ^2-3 R\right) \left(4 M+R^3 \omega ^2-2 R\right)}
\end{equation}
This leads to:
\begin{equation}\label{R1full}
 R_1= 2 \pi  \eta  R \omega ( 2 M-R)  \frac{m H(R)+n H'(R)}{p}
\end{equation}
where
\begin{eqnarray} 
m &=&8 M \left(3 M^2-6 M R+2 R^2\right)+R^4 \omega ^2 (5 R-19 M)\,, 
\nonumber
\\
n &=&4 R (2 M-R) \left(3 M^2+3 M R+R^4 \omega ^2-2 R^2\right) \,, 
\nonumber 
\\
p &=&\left(6 M+R^3 \omega ^2-3 R\right) \left(4 M+R^3 \omega ^2-2 R\right) [R (16 \pi  \eta  R \omega +i)-3 i M]\,.
\end{eqnarray}
We next provide the explicit forms of the partial derivatives of the unperturbed pressure with respect to the unperturbed density (i.e., the square of the sound speed, $c_s^2$) and with respect to the membrane radius $R$,
\begin{equation}
    c_s^2=\left(\frac{\partial p_0}{\partial \rho_0}\right)=-\frac{3 M^2-3 M R+R^2}{2 \left(6 M^2-5 M R+R^2\right)}\;,\;\;\;\left(\frac{\partial p_0}{\partial R}\right)=-\frac{f(R) \left(3 M^2-3 M R+R^2\right)}{8 \pi  R^2 (R-2 M)^2}
\end{equation}
Substituting \ref{rho1full}, \ref{p1full} and \ref{R1full} into \ref{EoSmp} and solving for $(H’/H)_R$, we obtain:
\begin{equation}
    \left(\frac{H'}{H}\right)_R=\frac{N}{D}
\end{equation}
where
\begin{align}
N &=-24 i M (R-2 M)^2 \left(6 M^2-6 M R+R^2\right)+16 \pi  R^8 \omega ^5 \left[3 M^2 (2 \zeta +9 \eta )-M R (5 \zeta +23 \eta )+R^2 (\zeta +5 \eta )\right]+\nonumber
\\
&\quad~i R^6 \omega ^4 \left[-195 M^3+256 \pi ^2 \zeta  \eta  (2 M-R) \left(30 M^2-31 M R+10 R^2\right)+221 M^2 R-82 M R^2+10 R^3\right]+\nonumber
\\
&\quad~16 \pi  R^4 \omega ^3 \left[M^4 (90 \eta -48 \zeta )-M^3 R (200 \zeta +171 \eta )+M^2 R^2 (320 \zeta +121 \eta )-2 M R^3 (72 \zeta +19 \eta )+5 R^4 (4 \zeta +\eta )\right]+\nonumber
\\
&\quad~4 i R^2 \omega ^2 \Big[M \left(54 M^4-216 M^3 R+234 M^2 R^2-96 M R^3+13 R^4\right)-512 \pi ^2 \zeta  \eta  (R-2 M)^2 (3 M^3-3 M^2 R+5 M R^2-\nonumber
\\
&\quad~3 R^3)\Big]-128 \pi  \omega  (2 M-R) \left[\eta  M \left(9 M^4-36 M^3 R+36 M^2 R^2-9 M R^3-R^4\right)+3 \zeta  R^2 (4 M-R) (R-2 M)^2\right]+\nonumber
\\
&\quad~256 i \pi ^2 \zeta  \eta  R^{10} \omega ^6 (2 M-R)\,, 
\nonumber
\\
D &=4 R (2 M-R) \Big\{6 i (R-2 M)^2 \left(3 M^2-3 M R+R^2\right)+16 \pi  R^4 \omega ^3 \Big[\zeta  (2 M-R) \left(3 M^2-7 M R+3 R^2\right)+\nonumber
\\
&\quad~\eta  (3 M-2 R) \left(3 M^2-6 M R+2 R^2\right)\Big]-i R^2 \omega ^2 \Big[27 M^4-27 M^3 R-256 \pi ^2 \zeta  \eta  (R-2 M)^2 \left(3 M^2+6 M R-2 R^2\right)+\nonumber
\\
&\quad~9 M^2 R^2-3 M R^3+R^4\Big]+16 \pi  \omega  (2 M-R) \left[\eta  \left(9 M^4-9 M^3 R-30 M^2 R^2+33 M R^3-8 R^4\right)-6 \zeta  R^2 (R-2 M)^2\right]\nonumber
\\
&\quad~+i R^6 \omega ^4 \left(256 \pi ^2 \zeta  \eta -1\right) (R-3 M) (R-2 M)\Big\}\,.
\end{align}
Finally, by setting $\eta_0=0$ and neglecting terms of order ${\cal O}(\omega^2)$, we recover \ref{BCH}.

\section{Extension to the membrane paradigm: $K^-_{ab}\propto K^+_{ab}$} \label{app:extension}
In this appendix we aim to extend the membrane paradigm by presenting a more general formulation for the extrinsic curvature within the membrane, a quantity that is typically (and arbitrarily) set to zero in the standard approach. Although any choice of $K^-_{ab}$ is possible, we will examine the case where $K^-_{ab}$ is proportional to $K^+_{ab}$,
\begin{equation}
   K^-_{ab}=A(r,\rho,p, \zeta,\eta)K^+_{ab}\;.
   \label{A1}
\end{equation}
Here, $A$ is a scalar function that depends on the radial coordinate and on the pressure, density, and viscosity of the membrane. We express $A$ perturbatively, restricting the expansion to first order in the perturbations, allowing us to study the background and the perturbed membrane separately,
\begin{equation}
    A(r,\rho,p, \zeta,\eta)=A_0(r_0,\rho_0,p_0)+\delta A(r,\rho,p,\zeta,\eta)\;.
    \label{jcA}
\end{equation}
 The first term in the expansion $A_0$ depends only on the background parameters, while $\delta A$ depends on the perturbed radial coordinate, density, pressure and on the membrane's response to external perturbations through the viscosity parameters.

By substituting \ref{A1} in \ref{condi}, we obtain the following junction conditions,
\begin{equation}
    [[h_{ab}]]=0,\qquad \left(1-A(r,\rho,p, \zeta,\eta)\right)\left[K^+h_{ab}-K^+_{ab}\right]=8\pi T_{ab}\;.
    \label{conA}
\end{equation}
Since we have adopted a more general form for the extrinsic curvature within the membrane, rather than fixing $K^-_{ab}=0$, this allows for new degrees of freedom in the system while keeping the number of constraints unchanged. At the background level, it is not possible to simultaneously express $\rho_0$, $p_0$ and $A_0$  solely in terms of the radial variable $r$. Therefore, we choose to express $p_0$ and $A_0$  as functions of both $\rho_0$ and $r$.
To this end, we calculate the trace of the second junction conditions in \ref{conA}, at zeroth order in the perturbations, and relate $A_0$ to the background pressure and density,
\begin{equation}
    A_0=\frac{R f'(R)+8 \pi  R \sqrt{f(R)} (\rho_0(R)-2 p_0(R))+4 f(R)}{r_0 f'(R)+4 f(R)}\,.
    \label{A0}
\end{equation}
Furthermore, from the diagonal terms of the junction conditions at zeroth order in the perturbations, we get
\begin{equation}
    p_0=-\frac{\rho_0 (2 f(R)+R f'(R))}{4 f(R)}\;.
    \label{p0A}
\end{equation}
By substituting \ref{p0A} in \ref{A0}, we can finally rewrite $A_0$ as,
\begin{equation}
  A_0 =1+\frac{4\pi R \rho_0}{\sqrt{f(R)}}\:.
  \label{Arho}
\end{equation}
At first-order in perturbation theory, we will treat the axial and polar sectors separately, following the same approach as in \ref{sec:membrane}. We can thus express the function $\delta A$
  as a sum of its axial and polar components, i.e. 
$\delta A=\delta A_{\rm ax}+\delta A_{\rm pol}$\;.
Moreover: (a) the trace of the shear tensor $\sigma_{ab}$ is zero in both the polar and axial sectors, which renders $T$ independent of $\eta$, (b) in the axial sector, the expansion term 
$\Theta$ vanishes, indicating that $T$ is independent of $\zeta$, (c) the perturbations $\delta R$, $\delta \rho$ and $\delta p$ are nonzero only in the polar sector. As a result of these considerations, we conclude that $\delta A_{\rm ax}=0$, a finding that can be confirmed by calculating the first-order perturbative trace of the junction conditions (\ref{conA}). Therefore, the function $\delta A$ is expressed as:
\begin{equation}
    \delta A=\delta A_{\rm pol}(r,\rho,p,\zeta)\;.
\end{equation}

For simplicity, we restrict our analysis to the $\ell=2$ axial case. Following the same steps as in \ref{sec:membrane}, we derive the boundary condition for the axial perturbation $h_0$.
Similarly  to the standard formalism of the membrane paradigm, for a generic choice of $\eta_0$, the ratio $(h_{0}'/h_{0})$  is given by \ref{BCaxetagen}, and it is independent of viscosity at zeroth order in frequency, $M\omega$. However, for $\eta_0=0$, the boundary condition for $h_0$ reads:
\begin{equation}
  \left(\frac{h_{0}'}{h_{0}}\right)_{R}=\frac{2 \left[8 \eta_1 M^2 (2 M-R)+\rho_0 R^4 \sqrt{f(R)}-3 M \rho_0 R^3 \sqrt{f(R)}\right]}{\rho_0 R^5 \sqrt{f(R)}-3 M \rho_0 R^4 \sqrt{f(R)}-8 \eta_1 M R (R-2 M)^2} + \mathcal{O}(M\omega)\,.
  \label{BCAaxial}
\end{equation}
From the boundary condition \ref{BCAaxial} we derive the magnetic TLN,
\begin{equation}
    k_{2}^{\rm M}=\frac{f(\epsilon,\rho_0,\eta_1)}{g(\epsilon,\rho_0,\eta_1)}\,,
\end{equation}
where
\begin{align}
f(\epsilon,\rho_0,\eta_1)&=M (\epsilon +1)^{7/2} (2 \epsilon -1) \rho_0-12 \eta_1 \epsilon ^{5/2}\,,
 \nonumber 
\\
g(\epsilon,\rho_0,\eta_1)&=-60 \eta_1 (2 \epsilon +1) \left[6 \epsilon  (\epsilon +1)-1\right] \sqrt{\epsilon } (\epsilon +1)^3+60 (\epsilon +1)^5 \log \left(\frac{\epsilon }{\epsilon +1}\right) \left[M (\epsilon +1)^{7/2} (2 \epsilon -1) \rho_0-12 \eta_1 \epsilon ^{5/2}\right]
   \nonumber
\\
&\quad+5 M (2 \epsilon -1) [2 \epsilon  (3 \epsilon  (2 \epsilon +7)+26)+25] (\epsilon +1)^{9/2} \rho_0\,.
\end{align}
This time, the TLN additionally depends on the background density, which is a function of $\epsilon$, $\rho_0=\rho_0(\epsilon)$. To understand the behavior of the TLN for $\epsilon \ll 1$, it is therefore essential to know how $\rho_0$ depends on the compactness parameter $\epsilon$. Assuming a simple power-law dependence, $\rho_0=c\cdot \epsilon^\alpha$, at leading order in $\epsilon$, we can distinguish three cases:
\begin{equation}
     k_{2}^{\rm M}=\begin{cases}
     0\,,  & \text{for } \alpha>1/2\\
       \frac{1}{-\frac{60\eta_1}{M c}+125+60  \log (\epsilon ) }\,,  & \text{for } \alpha=1/2\\
       \frac{1}{125+60 \log (\epsilon )}\,,& \text{for } \alpha<1/2
     \end{cases}
\end{equation}
We observe that when the density has a power-law form, the TLN depends on the viscosity $\eta_1$, at zeroth order in $\epsilon$, only when $\rho_0=c\cdot \sqrt{\epsilon}$. In the standard formalism of the membrane paradigm, the background density is given by \ref{rho0}, which corresponds to $\alpha=1/2$ and $c=-1/8 M \pi$.  Substituting these values of $\alpha$ and $c$ into \ref{Arho} confirms that $A_0=0$.


\end{widetext}
\bibliography{References}

\bibliographystyle{./utphys2}

\end{document}